\pdfoutput=1
\documentclass[a4paper,11pt]{article}

\usepackage{jheppub}

\usepackage{bm}  
\usepackage{amsmath}     
\usepackage{epsfig,epsf}
\usepackage{xcolor}
\usepackage{graphicx}
\usepackage{rotating}
\usepackage{lmodern}
\usepackage{lipsum}
\def\beq{\begin{equation}}
\def\eeq{\end{equation}}
\def\bea{\begin{eqnarray}}
\def\eea{\end{eqnarray}}

\def\eq#1{{Eq.~(\ref{#1})}}
\def\fig#1{{Fig.~\ref{#1}}}

\newcommand{\Lb}{\left(}
\newcommand{\Rb}{\right)}

\parskip=\itemsep               

\newcommand{\nn}{\nonumber}

\newcommand{\h}{\frac{1}{2}}

\newcommand{\ga}{\gamma}

\newcommand{\eval}[1]{\left \langle #1 \right \rangle}

\newcommand{\pom}{I\!\!P}
\newcommand{\intl}{\int\limits}
\newcommand{\Y}{\tilde{Y}}

\numberwithin{equation}{section}

\def\pom{{I\!\!P}}

\title{ High energy scattering in the Unitary Toy Model}
\author[a]{Alex Kovner,}
\author[b]{Eugene Levin,}
\author[c]{and Michael Lublinsky}
\affiliation[a]{Physics Department, University of Connecticut, 2152 Hillside Road, Storrs, CT 06269, USA}
\affiliation[b]{Department of Particle Physics, Tel Aviv University, Tel Aviv 69978, Israel}
\affiliation[c]{Physics Department, Ben-Gurion University of the Negev, Beer Sheva 84105, Israel}

\abstract{ We continue exploring the Unitary Toy Model (UTM)  as a 
playground for high energy collisions in QCD.  Our new approach is based on the diagonalization of the evolution Hamiltonian. Part of the spectrum can be identified with intercepts of dressed Pomerons.  Analogously to QCD, a 
multi-Pomeron expansion of the $S$-matrix is badly divergent asymptotic series. Yet we succeeded to establish resummation procedures resulting in a well behaved $S$-matrix. In addition the Hamiltonian possesses negative eigenvalues, which dominate the approach of the $S$-matrix to saturation. We are hopeful that important lessons about BFKL-based Pomeron calculus could be taken from the toy world to real QCD.
}

\begin{document}
\maketitle

\pagestyle{empty}
\newpage

\mbox{}

\pagestyle{plain}

\setcounter{page}{1}
\author{ }

\abstract{}
\dedicated{}
\preprint{}

\section{Introduction}

Zero transverse dimension toy models
\cite{ACJ,AAJ,JEN,ABMC,CLR,CIAF,MUSA,RS,BIT,SHXI,KOLEV,KLremark2,nestor,LEPRI,utm,utmm} can be viewed as a realization of the Pomeron calculus or more generally of the
Reggeon Field Theory (RFT). Over the years they have been intensively used to model high energy collisions in QCD. 
These models encode various fundamental features of QCD such as unitarity, but are much simpler than the latter and frequently 
solvable analytically.  Hence they provide a valuable playground to gain intuition about the dynamics of real QCD.

Our discussion focuses on two toy models. 
 The first one is the BFKL cascade model  which mimics  Mueller's dipole model in QCD \cite{MUDI} 
 (for  dipoles of  fixed sizes). We refer to this model as the BFKL cascade \cite{BFKL}, although it is a closer analog to the Balitsky-Kovchegov (BK) model in QCD \cite{BK} \footnote{The gluon emissions in the projectile wave function in BFKL and BK models are the same. The physical difference between the two is that BK equation \cite{BK} accounts for multiple scatterings on the target, while the original BFKL equation \cite{BFKL} does not.}.
 The BFKL cascade lacks both $s$ and $t$-channel unitarity \cite{KLremark2,utm} and hence has to be corrected. A simple
 modification of the BFKL cascade wich restores the proper unitarity properties is the Unitary Toy Model (UTM) first introduced in \cite{MUSA} and later explored in \cite{KLremark2,BIT,utm}. 
Additional improvements  to the model  have been proposed  (see \cite{utmm}), but they will not be addressed here. 

All of the  models  share the following simple probabilistic expression for 
 the total  $S$-matrix for scattering of $n$ dipoles of the projectile on $m$ dipoles of the target
  \beq \label{SMS}
 S(Y) \,\,=\,\,\sum_{n,m} e^{ - m \,n\,\gamma} \,P_n^P( Y_0)\,P_m^T( Y - Y_0)
 \eeq
 Here $\gamma \sim \alpha_s^2$ is the Born approximation to the low energy dipole-dipole scattering amplitude. 
 $P_n^P( Y_0) $ is the probability  to find $n$ dipoles  in the projectile boosted to rapidity $Y_0$, and similarly for the  target. The probabilities are found by solving the evolution equations in rapidity, and it is in the form of these equations that the models differ from each other.  The important consistency condition on the evolution is $t$-channel unitarity, which is essentially the requirement that  the $S$-matrix does not depend on  the reference frame, that is  the choice of $Y_0$ at  fixed $Y$\cite{MUSA,BIT,utmm}. UTM is defined to satisfy this condition.

Initial conditions for the evolution determine the dipole content of the colliding  particles. In this paper, we will be largely concerned  with dipole-dipole scattering, that is with initial condition $P_1^{P,T}(0)=1$, $P^{P,T}_{n>1}(0)=0$. We will also  briefly discuss DIS-like processes off a nucleus, in which the target is assumed to consist of many dipoles at rest.
 
While both the BFKL and UTM models have been extensively explored in the past, in this paper we address a somewhat different set of questions and use a new approach. Our formulation follows  algebraic approach based on diagonalization of the respective evolution Hamiltonians, for which 
we find spectrum and eigenfunctions. Part of the spectrum can be identified with Pomeron  
intercepts,  which differ between the models.  
Pomeron calculus  emerges as a representation of the $S$-matrix. 

The multi-Pomeron expansion in BFKL is an asymptotic, Borel summable series, while in UTM
it is  badly divergent. The UTM asymptotic series cannot be summed 
 via Borel resummation. Nevertheless the $S$-matrix in UTM  is physical and well defined, and we are able to establish resummation methods which circumvent the apparent divergence of the Pomeron calculus and yield the finite, physical result for the $S$-matrix.  
 
 A similar and a very old problem  also exists in QCD:  due to $1/N_c$ corrections, the $2n$-reggeon (reggeized gluon)  BKP states \cite{BKP} have intercepts  which grow faster  than  $n$ times the single BFKL Pomeron intercept.  As a consequence, for many years, this was believed to be the major challenge even to  very existence of the BFKL-based Pomeron calculus in real QCD. We are  hopeful that our explorations in the toy models provide an important insight into RFT in QCD, suggesting that the $1/N_c$ corrections  can be re-summed.

Beyond discussing the $S$-matrix we touch upon  the probability distributions.   The latter contain all the information about the dipole cascade and could be instrumental in understanding particle production in these toy models. This question is
postponed for a separate study.

We   review the BFKL cascade model in Sec. 2. UTM is introduced in Sec. 3.  
$S$-matrix in UTM is explored in Sec 4. Sec 5 is the summary section. Several Appendices complement 
our calculations.

 
 
   \section{The BFKL cascade }

 \subsection{The dipole-dipole scattering: probabilities and Pomeron exchanges}
    
  The simplest model describes a single BFKL-like cascade. It is specified by the following evolution 
equation  for the dipole probabilities
  $P^{\mbox{\tiny BFKL}}_n(Y) $: 
\beq \label{PBK}
\frac{ d\, P^{\mbox{\tiny BFKL}}_n(Y)}{ d\,Y} \,\,=\,\, - \Delta\,n\,P^{\mbox{\tiny BFKL}}_n(Y)\,\,+\,\,\Delta\,(n - 1)\,P^{\mbox{\tiny BFKL}}_{n-1}(Y)
\eeq
Here we regard the "BFKL intercept" parameter $\Delta$ as parametrically of order  $\Delta\sim\alpha_sN_c$ inspired by the QCD BFKL equation.

The solution for this equation for a single dipole initial condition is:
\beq \label{PNBK}
P_{n}^{BFKL}(Y)=\frac{1}{N(Y)}\left(1-\frac{1}{N(Y)}\right)^{n-1}\,\,\,\xrightarrow{Y\,\gg\,1/\Delta}\,\,\,\frac{1}{N(Y)}\exp\Lb - \frac{n\,-\,1}{N(Y)}\Rb
\eeq
where the  average dipole multiplicity in the BFKL cascade is
\beq\label{N}
 N(Y)\,=\, e^{ \Delta\,Y}
 \eeq
 The $S$-matrix calculated via eq.\eqref{SMS} with $ P^{\mbox{\tiny BFKL}}_n(Y)$ does not 
 depend on $Y_0$ only if we expand   the exponential factor $ e^{ - m \,n\,\gamma}$ to leading order in $\gamma$, 
 $ e^{ - m \,n\,\gamma} \approx 1-mn\gamma$:
 \begin{equation}\label{SBFKL}
 S^{\mbox{\tiny BFKL}}_{\mbox{\tiny LO}}=1-\gamma N(Y)
 \end{equation}
 
 Keeping all terms in the exponents leads to a nontrivial dependence of $S$ on $Y_0$. However, it turns out that 
 if both $Y-Y_0$ and $Y_0$ are large, the  dependence on $Y_0$ disappears in the  term that leads to the largest contribution.  Indeed,
  \begin{subequations}  
 \bea
\hspace{0cm} S^{\mbox{\tiny BFKL}}(Y) & =&\sum_{n,m} e^{ - m \,n\,\gamma} \,P^{\mbox{\tiny BFKL}}_n( Y_0)\,P^{\mbox{\tiny BFKL}}_m( Y - Y_0) \label{SMS4a}\\
&=&\sum_{k=0}^\infty 
 \frac{\Lb - \gamma\Rb^k}{k!}\underbrace{\Bigg( \sum_{n=0}^\infty  n^k \,P_n( Y_0) \Bigg)}_{c_k\Lb Y_0\Rb}\underbrace{ \Bigg( \sum_{m=0}^\infty  m^k \,P_m( Y - Y_0) \Bigg)}_{c_k\Lb Y - Y_0\Rb}\,\, \label{SMS4b}
 \eea
  \end{subequations}  
\eq{SMS4b} gives the expression for the scattering amplitude which can be viewed as the sum of multi Pomeron exchanges. Each of these exchanges is proportional to $\Lb e^{\Delta Y} \Rb^n$. \eq{SBFKL} is the contribution of the first two terms in \eq{SMS4a}. 
Lets define short notations $N_1=N\Lb Y_0\Rb$, $N_2= N\Lb Y  - Y_0\Rb$, and assume both to be large.
We start with \eq{SMS4a} which takes the form
   \begin{subequations}    \bea 
 S^{\mbox{\tiny BFKL}}(Y) & =&\sum_{n,m} e^{ - m \,n\,\gamma} \,P^{\mbox{\tiny BFKL}}_n( Y_0)\,P^{\mbox{\tiny BFKL}}_m( Y - Y_0) \,\simeq\,
 \sum^{\infty}_{n=1,m=1} e^{ - m \,n\,\gamma} \frac{e^{- \frac{n}{N_1}}}{N_1}\, \,\frac{e^{- \frac{n}{N_2}}}{N_2} \label{ SMS41a} \\
\simeq\,(\gamma \,\ll\,1) &\simeq& \frac{1}{N_1\,N_2} \intl^\infty_0 d \bar{m}\intl^\infty_0 d \bar{n}\exp\Lb - \gamma\bar{n}\bar{m}- \gamma\bar{n}-\gamma\bar{m}    - \gamma\,- \bar{n}/N_1 - \bar{m}/N_2\Rb  \label{ SMS41b} \\ 
 &=& \frac{1}{\gamma N_1 N_2}e^{\frac{1}{\gamma  \text{N}_1 \text{N}_2}+\frac{1}{\text{N}_1}+\frac{1}{\text{N}_2}} \Gamma \left(0,\frac{(\gamma\,\text{N}_1 \, +\,1) ( \gamma\,\text{N}_2 +1)}{\gamma\,\text{N}_1 \text{N}_2  }\right)  \label{SMS41c}\\
\xrightarrow{N_2\gg1,N_1\gg1} &  &\,\frac{1}{\gamma\,N_1\,N_2} \Bigg( - \ln \gamma + \mathcal{O}(1/N_1) \,+  \mathcal{O}(1/N_2)\Bigg)  \label{SMS41d}\eea
  \end{subequations}   
 where $\bar{n} = n -1$, $\bar{m} = m - 1$,  and $\Gamma\Lb 0, z\Rb$ is  the incomplete Gamma function   (see formulae {\bf 8.35} of Ref.\cite{RY}). The leading term does not depend on $Y_0$. However, all other terms do depend on the value of $Y_0$.

 It is instructive to see how  the result \eq{SMS41d} arises from the "Pomeron calculus", i.e. \eq{SMS4b} which represents the 
 scattering amplitude as a sum of  Pomeron exchanges.
 The power moments $c_k$  are given by
 \beq \label{SMSM1}
 c_k\Lb N \Rb= \frac{1}{N}\sum_{n=1}^{\infty}  n^k \exp\Lb - \frac{n}{N}\Rb\,=\,\,\text{Li}_{-k}\left(e^{-1/N}\right)\,\,\xrightarrow{ N\gg 1}  k! N^k \,\,+\,\,\frac{\zeta\Lb - k\Rb}{N} \,\,+\,\,\mathcal{O}\Lb 1/N^2\Rb
 \eeq 
 where $\zeta(z)$ is the Riemann zeta function. For $\zeta\Lb - k\Rb$ we use  the following equation (Abel–Plana formula):
 \beq \label{SMSM2}
 \zeta(-k) = - \frac{1}{k + 1}  + \tilde{\zeta}\Lb -k\Rb;~~\mbox{with}~~
 \tilde{\zeta}\Lb -k\Rb\,\,=\,\,\h \,+\,2 \intl^\infty_0 d t \frac{\sin\Lb k \arctan(t)\Rb \Lb 1 - t^2\Rb^{k/2} }{ e^{ 2 \pi t} \,-\,1}
 \eeq 
$\tilde{\zeta}\Lb k\Rb $  is equal $\h + $ function that decreases faster than $1/k$. The first term in \eq{SMSM2}  has a simple meaning if in \eq{SMSM1} summation is replaced by 
integration
  \beq \label{SMSM3}
 c_k\Lb N \Rb= \frac{1}{N}\int_{n=1}^{\infty} dn\, n^k \exp\Lb - \frac{n}{N}\Rb\,=
 k! \,N^k \,\,-\,\,\frac{1}{k + 1} \frac{1}{N} \,\,+\,\,\mathcal{O}\Lb \frac{1}{N^2}\Rb
 \eeq.
Substituting $ \zeta(-k) = - \frac{1}{k + 1}$ into \eq{SMSM1}, for the $S$-matrix  we obtain
 \bea \label{SMSM4}
  \hspace{0cm}&&S^{\mbox{\tiny BFKL}}(Y) 
=\sum_{k=0}^\infty 
 \frac{\Lb - \gamma\Rb^k}{k!}\,c_k\Lb Y_0\Rb\, c_k\Lb Y-Y_0\Rb\,\simeq\,
 \sum_{k=0}^\infty  \frac{\Lb - \gamma\Rb^k}{k!}\Lb k!\,N_1^k\ \,- \frac{1}{k+1} \frac{1}{N_1} \Rb\, \Lb k!\, N^k_2 \,- \frac{1}{k+1} \frac{1}{N_2} \Rb \nn\\
  && \simeq\,\,\frac{1}{\gamma \,N_1\, N_2}   e^{\frac{1}{\gamma  N_1\, N_2}} E_1\left(\frac{1}{ \gamma N_1\,N_2}\right) \,\, -\,\,\frac{\ln\Lb 1 \,\,+\,\,\gamma N_1\Rb}{
\gamma\,N_1\,N_2} \,\,-\,\, \frac{\ln\Lb 1 \,\,+\,\,\gamma N_2\Rb}{
\gamma\,N_1\,N_2}\nn\\
&&\xrightarrow{ \gamma N_1;\,\gamma \, N_2 \gg\,1} \,
\frac{1}{\gamma \,N_1 \,N_2} \Bigg\{ \ln\Lb \gamma N\Lb Y\Rb\Rb\,-\,
\ln\Lb \gamma N_1\Rb\,-\,
\ln\Lb \gamma N_2\Rb\Bigg\} \,=\,\frac{ \ln \Lb 1/\gamma\Rb}{\gamma \,N_1\,N_2}
\,\,=\, \frac{ \ln \Lb 1/\gamma\Rb}{\gamma}\,e^{-\Delta\,Y}
\eea
where $E_1\Lb z \Rb = - Ei\Lb - z\Rb$ is the Exponential integral (Ref.\cite{RY}, formulae {\bf 8.21}).
 \eq{SMSM4} gives the same asymptotics as \eq{SMS41d}. 
 
 In the context of this calculation we note that in order to obtain the correct asymptotics using the moment expansion (or "Pomeron calculus") we need to keep the subleading term in \eqref{SMSM3} for large $N$. If we were to drop this subleading term we would get the factor $\ln(\gamma N(Y))$ instead of $\ln 1/\gamma$ in front of the exponential in \eqref{SMSM4}, i.e. a significantly different asymptotics at large rapidity. This may sound surprising, since this term is a very small correction to high moments at any $Y$, and to all moments at large $Y$.  However while the series representation in terms of probabilities is convergent, as probabilities for large number of dipoles $n$ always decrease as $n\rightarrow\infty$, the expansion in moments (or in Pomeron exchanges) is a very different animal. The individual terms in this expansion grow factorially with $k$ and exponentially in $Y$. Thus even a small correction to a fixed moment can generate a leading correction to the whole sum. This should caution us against using naive Pomeron calculus (i.e. approximating $n$-Pomeron exchange by a simple product of $n$ single Pomerons) in a frame where both $Y_0$ and $Y-Y_0$ are large, and thus a single term in the Pomeron series is given by a product of two large numbers, $c_k(Y_0)c_k(Y-Y_0)$ as in \eqref{SMS4b}.
 Thankfully,  terms suppressed by additional powers of $1/N$ in \eqref{SMSM3} do not contribute to the asymtotic result.

We emphasize that in the BFKL cascade model, the $Y_0$ independence is approximate only
and is seen in a frame when both projectile and target are boosted to sufficiently high energies. The resulting
expressions for the $S$-matrix also differ significantly from the result computed in the target rest frame.
The latter we quote below in \eq{SA1b}. While the exponential fall with the total rapidity is the same, the leading 
coefficient differs by $-\ln\gamma$, which is quite dramatic. This signifies the problem of the BFKL cascade model
and its frame dependence. 

\subsection{The $S$-matrix in the target rest frame}
Let us now consider the calculation of the scattering matrix in the rest frame of the target. This means that we evolve only the projectile probability distribution, while the target remains fixed. In most of this paper we are interested in the target comprised of a single  dipole.

To calculate the  S-matrix it is convenient to introduce the notion of the generating functional.
 
 \subsubsection{Generating functional and eigenfunctions}
 
 The standard definition of the generating functional\cite{MUDI,LELU} is
\beq
Z^{BFKL}(u, \Y)\equiv \sum_{n} u^n\, P_n^{BFKL}(\Y)
\eeq
Beyond the formal meaning of generating the probabilities $P_n$ in the projectile, the functional $Z(u)$ for $u=e^{-\gamma m}$ also has a direct physical meaning of a scattering matrix of the projectile on a target consisting of  $m$ dipoles, calculated in the target rest frame.
The BFKL equation for the generating function can be written as
\beq\label{Zbfkl}
\frac{\partial}{\partial \tilde Y}Z^{BFKL}(u)\,=\,{\cal H}_{BFKL}\,Z^{BFKL}(u)
\eeq
where $\tilde Y=\frac{\Delta}{\gamma}Y$ 
and the BFKL Hamiltonian operator is
\beq \label{lbfkl}
 {\cal H}_{BFKL}= \gamma u(u-1)\frac{d}{du}
\eeq
 The standard way of solving an evolution equation of the type \eqref{Zbfkl} is first to find the 
 eigenfunctions of the evolution operator.
The operator $ {\cal H}_{BFKL}$ is not Hermitian and it therefore has distinct right and left eigenfunctions.
It is straightforward to verify that the following are its right eigenfunctions:
\beq
\Phi_n^{BFKL}\Lb u\Rb\,=\, \left({u-1\over u}\right)^n=\phi^n\,
~~~~~~~~~~~~~~~~~
\Psi_n^{BFKL}\Lb u\Rb\,=\, \left({u-1\over u}\right)^{-n}=\phi^{-n}\,
\eeq
with the corresponding eigenvalues $\Delta^{BFKL}_n=n\gamma$ and $\tilde \Delta^{BFKL}_n=-n\gamma$.
Here we have defined for convenience
\begin{equation}
\phi\equiv \frac{u-1}{u}
\end{equation}
Note, that in principle $n$ in the above expressions is not quantized, and therefore in this sense the spectrum is continuous. However our goal will be to expand the generating function in these eigenfunctions. By definition $Z(u)$ should be expandable in powers of $u$, which would suggest that the relevant eigenfunctions for the expansion are $\Psi_{BFKL}^n$ for integer $n$. On the other hand for $u$ close to unity we expect $Z(u)$ to have a good Taylor expansion in powers of  $1-u$. This singles out the set $\Phi_{BFKL}^n$ for integer $n$.
We will therefore only need the eigenfunctions $\Phi_{BFKL}^n$ and $\Psi_{BFKL}^n$ for integer values of $n$.

The left eigenfunctions with the same eigenvalues are
\begin{eqnarray}\label{left1}
 \tilde\Phi_{n}^{BFKL}\Lb u\Rb\,=\, {1\over (1-u)} \,\left({u-1\over u}\right)^n\,,~~~~~~~~~~~
 \tilde\Psi_{n}^{BFKL}\Lb u\Rb\,=\, {1\over (1-u)} \,\left({u\over u-1}\right)^n
\end{eqnarray}

\begin{figure}[ht]
    \centering
  \leavevmode
      \includegraphics[width=10cm]{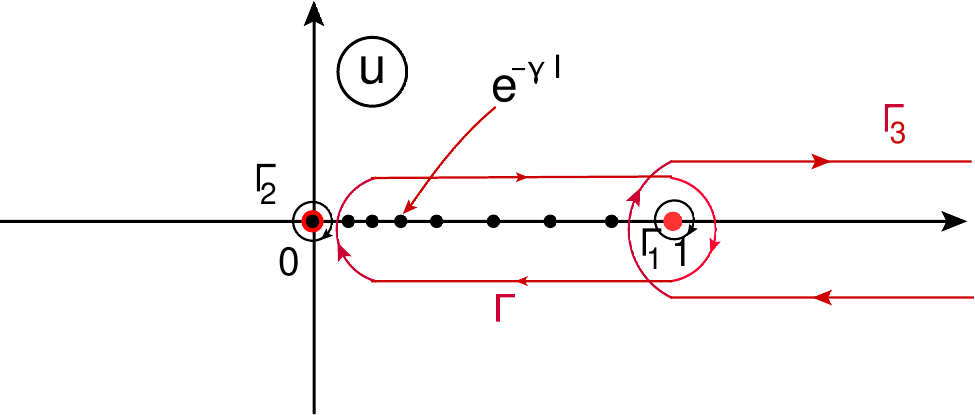}  
      \caption{ Contours of integrations in the complex $u$ plane. The red dots mark the poles in the BFKL cascade model while the black dotes are those in the UTM.}
\label{contga}
   \end{figure}

Since ${\cal H}_{BFKL}$ is not Hermitian, it's left and right eigenfunctions are different. In general for finite dimensional spaces there is a very useful theorem, that states that the left and right eigenfunctions $\phi^L_n$ and $\phi^R_m$ that correspond to different eigenvalues are orthogonal to each other. Therefore with appropriate normalization one can use the set of all right eigenfunctions to "invert" the left eigenfunctions
\beq\label{or}
\langle \phi^L_n|\phi^R_m\rangle=\delta_{nm}
\eeq
This property extends to infinite dimensional spaces if the eigenfunctions in question are well behaved (normalizable). Our eigenfunctions on the other hand clearly are not normalizable as they have nonintegrable singularity either at $u=1$ or $u=0$. Nevertheless, one can show by direct calculation that with an appropriate definition of scalar product \eqref{or} still holds. In particular
the two sets of eigenfunctions are orthonormal with the scalar product defined as a contour integral in the complex plane:
\beq\label{ort11}
\frac{1}{2\pi i}\oint_{\Gamma_1} {du\over u} \,\,\tilde\Psi_{m}^{BFKL}\Lb u\Rb\, \Phi_n^{BFKL}\Lb u\Rb\,
=\,\frac{1}{2\pi i}\oint_{-\Gamma_2} {du\over u} \,\,\tilde\Psi_{m}^{BFKL}\Lb u\Rb\, \Phi_n^{BFKL}\Lb u\Rb
\,=\, \delta_{mn}
\eeq
where the contour $\Gamma_1$ in the complex plane is taken to encircle {\bf only} the point $u=1$ while
 $\Gamma_2$ is the contour that {\bf only} encircles the point $u=0$ (see Fig. \ref{contga}). Equivalently,
 \beq\label{ort21}
\frac{1}{2\pi i}\oint_{\Gamma_1} {du\over u} \,\,\Psi_{m}^{BFKL}\Lb u\Rb\, \tilde\Phi_n^{BFKL}\Lb u\Rb\,
=\,\frac{1}{2\pi i}\oint_{-\Gamma_2} {du\over u} \,\,\Psi_{m}^{BFKL}\Lb u\Rb\, \tilde\Phi_n^{BFKL}\Lb u\Rb
\,=\, \delta_{mn}
\eeq
In the following sections we will see that a similar scalar product that preserves the property \eqref{or} can also be defined in the UTM.

As the previous equation suggests we can choose either $\Gamma_1$ or $\Gamma_2$ as the contour for the definition of the scalar product. 

When expanding a function $Z^{BFKL}(u)$ in the basis of right eigenfunctions, in order to extract the coefficients of the expansion we can use the "operator" representation of the left eigenfunctions:
\begin{eqnarray}\label{left2}
&& \tilde\Phi_{n}^{BFKL}\Lb u\Rb\,=\,\,\delta(u)\,\, \frac{1}{n!}\partial_{\phi^{-1}}^n=(-1)^n\delta(u)\frac{1}{n!}\left((u-1)^2\frac{d}{du}\right)^n
,~~~~~~~~~~~\nonumber\\
&& \tilde\Psi_{n}^{BFKL}\Lb u\Rb\,=\, \delta(1-u) \,\,\frac{1}{n!}\partial_\phi^{n}=\delta(1-u)\frac{1}{n!}\left(u^2\frac{d}{du}\right)^n
\end{eqnarray}
This corresponds to the choice of $\Gamma_1$ in eq.\eqref{ort11} and $\Gamma_2$ in eq.\eqref{ort21}.

The problem of solving the BFKL equation now becomes similar to a Schroedinger equation, except with the Hamiltonian which is not Hermitian. Nevertheless, we can expand the generating function and therefore the scattering matrix in the set of right eigenfunctions. Here we can choose either $\Phi_n$ or $\Psi_n$ as the basis for expansion, since each one of them forms a complete basis (a peculiarity of a non Hermitian Hamiltonian).
Which set to choose is dictated by convenience. 

Suppose first we are in the genuine BFKL situation, where the target is dilute. This means that the single dipole scattering amplitude is small, or $\gamma\ll 1$, or $1-u\ll 1$. In this situation it is clear  that the generating functional should be expanded in Taylor series in $1-u$.  Since in this regime $\Phi_n^{BFKL}\approx (u-1)^n$, it means that we should be expanding the generating function in the basis of $\Phi_n^{BFKL}$:
\beq \label{bfkl1}
Z^{BFKL}\Lb u,  \Y \Rb\,\,=\,\,\sum^\infty_{n=0} C_n\,\,\Phi_n^{BFKL}\Lb u\Rb\, e^{ n\,\gamma\,\Y} 
 \,. \eeq
We can use $\tilde\Psi_{n}^{BFKL}$ to find the expansion coefficients $C_n$.
For this we could either use directly the representation \eqref{left2}, or use \eqref{left1} noting that
\beq \label{ZC}
\frac{1}{2\pi i}\oint_{\Gamma_1} {du\over u} \,\,\tilde\Psi_{m}^{BFKL}\Lb u\Rb\, 
Z^{BFKL}\Lb u,  \Y \Rb\,=\, C_m\, e^{m\,\gamma\,\Y}
\eeq
Applying to a single dipole initial condition, $Z^{BFKL}_{(1)}\Lb u,  \Y\,=\,0 \Rb\,=\,u$ we find\footnote{Interestingly, if we were to try to use the contour $\Gamma_2$  to calculate the coefficients $C_m$ we  would find that they all vanish. This is the consequence of the fact that the expansion \eqref{bfkl1} converges at $u=1$ with the unit radius of converges, while the contour $\Gamma_2$ includes a region $u<0$ and is therefore outside of the radius of convergence. },
\beq\label{cm1}
C_m^{(1)}\,=\frac{1}{2\pi i}\,\oint_{\Gamma_1} {du\over u}\,\, \tilde\Psi_{m}^{BFKL}\Lb u\Rb\,  u \,=\,1
\eeq
and
\beq\label{zbfkl}
Z^{BFKL}_{(1)}\Lb u,  \Y \Rb\,\,=\,\,
 \sum^\infty_{n=0} \left({u-1\over u}\,e^{\gamma\,\Y}\right)^n
\eeq

The $S$-matrix of a single dipole in the target rest frame is then
\beq \label{SA1b}
S\Lb \Y\Rb\,\,=\,\,Z^{BFKL}_{(1)}\Lb e^{- \gamma},\Y\Rb \,\,=\,\, \sum^\infty_{n=0} \left({(1-e^{\gamma})}\,e^{\gamma\,\tilde Y}\right)^n=\frac{1}{1-(1-e^\gamma)e^{\gamma\,\tilde Y}}
\eeq
which reproduces the well known solution of the BK equation.

More generally for an $n$ dipole initial state,  $Z^{BFKL}_{(n)}\Lb u,  \Y\,=\,0 \Rb\,=\,u^n$. we
have
\beq
C^{(n)}_m\,=\frac{1}{2\pi i}\,\oint_{\Gamma_1} {du\over u}\,\, \tilde\Psi_{m}^{BFKL}\Lb u\Rb\,  u^n \,=\,\frac{(m+n-1)!}{m!\,(n-1)!}
\eeq
\beq
Z^{BFKL}_{(n)}\Lb u,  \Y \Rb\,\,=\,\,
 \sum^\infty_{m=0}\frac{(m+n-1)!}{m!(n-1)!} \left({u-1\over u}\,e^{\gamma\,\tilde Y}\right)^m
\eeq
And
\beq \label{SA1n}
S^{(n)}\Lb \Y\Rb\,\,=\,\,Z^{BFKL}_{(n)}\Lb e^{- \gamma},\Y\Rb \,\,=\,\, \sum^\infty_{m=0} \frac{(m+n-1)!}{m!(n-1)!}
\left({(1-e^{\gamma})}\,e^{\tilde Y}\right)^m=\left[ \sum^\infty_{m=0} \left({(1-e^{\gamma})}\,e^{\gamma\,\tilde Y}\right)^m\right]^n
\eeq

Now consider the opposite situation, i.e. dense target.  This means that $\gamma\gg 1$, and $u\ll 1$. This is not what we normally call BFKL, but this is certainly in the region of applicability of the BK equation which applies to scattering of a small number of dipoles on a dense target. Now one would  want to expand $Z(u)$ in powers of $u$ rather than in powers of $1-u$, and therefore use the $\Psi^{BFKL}_n$ basis.
\beq \label{bfkl2}
Z^{BFKL}\Lb u,  \Y \Rb\,\,=\,\,\sum^\infty_{n=0} B_n\,\,\Psi_n^{BFKL}\Lb u\Rb\, e^{ -n\,\gamma\,\Y} 
 \,. \eeq
The coefficients can be found in the analogous way{\footnote{Here the situation is the mirror image of \eqref{cm1}. If we try to use the contour $\Gamma_1$ to extract the expansion coefficients we will fail. This is because the expansion \eqref{bfkl2} converges around $u=0$ and not around $u=1$, and the contour $\Gamma_1$ is partially outside the radius of convergence.}:
\beq \label{ZC}
\frac{1}{2\pi i}\oint_{-\Gamma_2} {du\over {u}} \,\,\tilde\Phi_{m}^{BFKL}\Lb u\Rb\, 
Z^{BFKL}\Lb u,  \Y \Rb\,=\, B_m\, e^{-m\,\gamma\,\Y}
\eeq
For a single dipole initial condition we find
\beq
B_m^{(1)}=-1, \ \ m>0
\eeq
Then
\beq\label{SA1c}
Z^{BFKL}_{(1)}\Lb u,  \Y \Rb\,=-\sum^\infty_{m=1}\left(\frac{u}{u-1}e^{-\gamma\,\tilde Y}\right)^m=1-\frac{1}{1+\frac{u}{1-u}e^{-\gamma\,\tilde Y}}=\frac{1}{1-(1-\frac{1}{u}) e^{\gamma\,\tilde Y}}
\eeq
which unsurprisingly coincides with \eqref{SA1b}.

For an $n$ dipole initial condition, 
$B^{(n)}_{m< n}=0$, while
\beq
B^{(n)}_{m\ge n}=\frac{1}{2\pi i}\,\oint_{-\Gamma_2} {du\over {u}}\, \tilde\Phi_{m}^{BFKL}\Lb u\Rb\,  u^n = \sum_{k=1}^n(-1)^k\frac{n!(k+m-1)!}{k!(n-k)!(k-1)!m!}\,=\,
(-1)^n\frac{(m-1)!}{(m-n)!\,(n-1)!}\,=\,-\,C^{(n)}_{-m}
\eeq

It is  instructive to compare the two expansions \eqref{SA1b} and \eqref{SA1c}, now both considered for arbitrary $\gamma$. For small $\gamma\ll 1$, \eqref{SA1b} is a good expansion, i.e. has decreasing terms for initial rapidity, and up to "critical" rapidity $\tilde Y_C\sim \frac{1}{\gamma}\ln\frac{1}{\gamma}$. For higher rapidities the individual terms are growing and to make sense out of it, the whole series has to be resummed. For large $\gamma\gg 1$ the expansion is poor right from the very beginning, and has to be resummed even at initial rapidity. As for \eqref{SA1c}, this for large $\gamma\gg 1$ is good at any rapidity, since for high rapidities the individual terms only get smaller. For small $\gamma\ll 1$ on the other hand, the expansion is poor at initial rapidity and all the way up to $\tilde Y_C$. For higher rapidities, $\tilde Y>\tilde Y_C$ however the expansion becomes good and gives reliable results only keeping a small number of terms.

Again we stress for future reference, that the expansion eq.\eqref{SA1b} for $Y<Y_c$ is convergent and not asymptotic - all the terms in the series are decreasing.

A comment is due here about the origin of the two sets of eigenvalues and the corresponding eigenfunctions. The positive eigenvalues $n\gamma$ are the zero dimensional analogs of the familiar n-Pomeron intercepts of QCD. These lead to the exponential growth of the scattering amplitudes at weak coupling and small rapidity. We will discuss the Pomeron calculus in more detail in the next section. On the other hand we are not so familiar with the negative eigenvalues $-n\gamma$ in QCD. The eigenfunctions corresponding to those are singular at $u=1$ and therefore, as discussed above is not a good basis for expansion at small rapidity. However for large rapidity where the scattering matrix is small, these eigenfunctions dominate the $s$-matrix, as is obvious from \eqref{SA1c}. Thus this set of eigenvalues governs the approach of the scattering matrix to saturation. In the context of QCD the approach to saturation in BK evolution is given by the Levin-Tuchin law \cite{LETU}, which indeed yields the $s$-matrix exponentially decreasing with rapidity. We thus identify the largest negative eigenvalue with the zero dimensional incarnation of the Levin-Tuchin decay. The rest of the negative eigenvalues provide correction to this leading exponential decay, and presumably should exist in QCD as well as subleading corrections close to saturation.




 \section{The UTM} 
 
 The UTM  was proposed  in Ref.\cite{MUSA} (see also Refs.\cite{KLremark2,BIT,utm}) as the model that satisfies  the $t$ channel unitarity condition. It was shown in \cite{utm} that it is also $s$-channel unitary.
 Requiring $Y_0$ independence of the $S$-matrix in \eq{SMS} one is lead to the evolution equation for  $P^{\mbox{\tiny  UTM}}_n\Lb Y\Rb$ 
   \cite{MUSA,KLremark2,BIT}:
 \beq \label{MEQ}
\frac{d P^{\mbox{\tiny  UTM}}_n(Y)}{ d Y}\,=\,- \frac{\Delta}{\tilde{\Delta}_1} \Lb 1\,-\,e^{- \gamma n}\Rb P^{\mbox{\tiny  UTM}}_n(Y) \,\,+\,\,
\frac{\Delta}{\tilde{\Delta}_1} \Lb 1\,-\,e^{- \gamma(n - 1) }\Rb\,P^{\mbox{\tiny  UTM}}_{n-1}(Y)
\eeq
 with $\tilde{\Delta}_1 = 1 - e^{- \gamma}\approx \gamma$ at weak coupling. While $\tilde\Delta_1$ is a free parameter of the UTM model,
 the specific choice is made so to exactly match the single Pomeron intercept of the BFKL cascade model (see below).

The crucial difference between (\eq{MEQ}) and  the BFKL cascade (\eq{PBK}) is that in the UTM the dipole emission probability saturates when the   number of dipoles is large $n>1/\gamma$
\footnote{in Ref. \cite{utmm} we have proposed an even more realistic model in which multiple dipole emissions are allowed in a single step of the evolution, but we are not considering it in this paper.}.

\begin{boldmath}
\subsection{The generating functional}
\end{boldmath}

Again we define the generating functional as
\beq
Z(u, \Y)\equiv \sum_{n} u^n\, P_n(\Y).
\eeq
where $\Y = \frac{\Delta}{\tilde\Delta_1} \,Y$. Notice that $\Y$ in the UTM is slightly different from that in the BFKL cascade.
In the UTM, the generating functional obeys the equation (see Eq. (2.16) of Ref.\cite{utmm}):
\beq\label{ZEQG}
\frac{\partial}{\partial \Y}Z(u, \Y)=-(1-u)\left(1-e^{-\gamma u\frac{\partial}{\partial u}}\right)Z(u,  \ Y)=\,( u - 1) \Bigg( Z(u,  \Y) - \,Z\Lb e^{-\gamma}\,u,\,  \Y \Rb\Bigg)
\eeq
with the initial and boundary conditions for a single dipole projectile:
  \bea
&& \mbox{Initial condition:} ~~Z\Lb u,  \Y =0\Rb\,=\,u
\\
&& \mbox{Boundary  condition:}~~ Z\Lb u=1,  \Y \Rb\,\,=\,\,1.
  \label{BC}    \eea
  The second condition \eqref{BC} can be also imposed as the initial condition at $\Y=0$, since it is preserved by the evolution \eqref{ZEQG} at any $\Y$.
     As in the BFKL case, the evolution of the generating function is generated by a non-Hermitian operator, which for the UTM is
\beq\label{L}
{\cal H}_{UTM}\,=\,-\,(1-u)\,\left(1-e^{-\gamma u\frac{\partial}{\partial u}}\right)
 \eeq

  \subsection{The UTM Pomeron calculus  }


 As is obvious from the previous discussion, the UTM is a well defined, t and s - channel unitary model that yields a finite $S$-matrix which at large rapidity falls to zero. It also exhibits saturation, as the dipole emission probability (per unit rapidity) for large dipole number $n$ approaches a constant as $n\rightarrow\infty$. Nonetheless, if we were to approach the model from the point of view of Pomeron calculus based on the BFKL approximation, we would have to face a serious problem.
 
 The BFKL Pomeron calculus is understood as perturbation theory around the BFKL limit of the Hamiltonian. To facilitate this we rewrite the UTM Hamiltonian as
  \beq \label{TAK4}
\mathcal{H}_{\rm UTM}\,\,=\,\, -\,\,
 t\,\Lb1-e^{\gamma (1 - t)\frac{\partial}{\partial t}}\Rb\,\,=\,\,
 \underbrace{\tilde\Delta_1\, t \frac{\partial}{\partial t}}_{ \mathcal{H}_0} \,-\,  
 \underbrace{ t\,\Lb1-e^{\gamma (1 - t)\,\frac{\partial}{\partial t}}\,\,+\,\,\tilde\Delta_1\,\frac{\partial}{\partial\,t}\Rb}_{ \mathcal{H}_I} 
 \eeq 
 where $t= 1-u$.
Using the variable $t$ rather than $u$ is natural at weak coupling/small rapidities as it has the meaning of the dipole-dipole scattering amplitude and is small in this regime. 

The choice of $\mathcal{H}_0$ dictated by 
 the interpretation of the zero dimension models as the QCD with fixed dipole size \cite{BFKL,KOLEB}. 
 At zeroth order in the perturbation, the eigenfunctions of $\mathcal {H}_{UTM}$ are pure powers
 \begin{equation}\label{tn}
 \mathcal{H}_0\,t^n\,=\,\Delta^0_n\,t^n; \ \ ~~~ \ \ \Delta^0_n\,=\,\tilde\Delta_1\, n\,.
 \end{equation}
 Although \eqref{tn} is valid for arbitrary $n$ in the context of the Pomeron calculus we choose to consider positive integer values of $n$ only. This choice corresponds to summation of BFKL ladders in QCD, and  ensures that expansion of the scattering matrix in Taylor series in $t$ is interpretable as expansion in multi Pomeron exchanges.
 In the same spirit we identify the perturbative BFKL Pomeron with the first eigenfunction 
 $\pom(\tilde Y)\equiv t\,e^{{\tilde\Delta_1} \Y}\,=\,t\,e^{{\Delta} Y}$.
 
 We can write the Hamiltonian in terms of  the "Schroedinger picture" Pomeron field  $\pom(\tilde Y=0)\equiv t$. In addition to $\mathcal{H}_0$, the Hamiltonian contains $n\rightarrow m$ Pomeron vertices
  \begin{eqnarray}\label{pomc}
 \mathcal{H}_{\rm UTM}&=& \tilde\Delta_1\, t\frac{\partial}{\partial t}-\sum_{m=1}^\infty\sum_{n=m-1}^{\infty}(-1)^{m-1}\left[1-\delta_{m=1,n=0}-\delta_{m=1,n=1}\right]\frac{1}{(m-1)!(n-m+1)!}\tilde{\Delta}_1^nt^m\left(\frac{\partial}{\partial t}\right)^n\nonumber\\
 &=&\tilde{\Delta}_1\,t\frac{\partial}{\partial t}-\sum_{m,n}^\infty V_m^nt^m\left(\frac{\partial}{\partial t}\right)^n
 \end{eqnarray}
  with
  \beq
  V_m^n=(-1)^{m-1}\frac{1}{(m-1)!(n-m+1)!}\tilde\Delta_1^n; \ \ \ \ \ n\ge m-1; \ \ \ \ m\ge 1
  \eeq
  except $(m=1,n=0)$ and $(m=1, n=1)$. 
  
The interaction contains  vertices  $ \pom^n \to  \pom^m$ for  an infinite number of values of $n$ and $m$ even though in terms of the number of dipoles, the UTM Hamiltonian only allows emission of a single dipole in one step of the evolution (see \eq{MEQ}).
The UTM
is therefore an example of a simple Pomeron calculus with the infinite number of vertices, which describes a fully unitary system with saturation. 

There are three distinct types of Pomeron vertices in \eqref{pomc}. The first type is $V_m^n$ with $n=m-1$. These are the Pomeron splittings. All the splitting vertices increase the number of bare Pomerons by exactly one. The second type is $V_m^n$ with $n>m$. Those are the Pomeron merging vertices. In UTM any number of Pomerons can disappear as a result of a merging via a single merging vertex. Finally the third type are vertices with $n=m$. Those preserve the number of Pomerons, but give corrections directly to the intercept of the $n$-Pomeron state (see \fig{dia}).

Imagine now a perturbative calculation of an $n$-Pomeron intercept in this Pomeron calculus. In the language of the previous section that corresponds to calculating  perturbatively eigenvalues of $\mathcal{H}_{UTM}$. It is the common lore that the first most important contribution in such a calculation comes from the so called "enhanced diagrams", i.e. diagrams with small Pomeron loops. There are many types of such diagrams in the UTM, since there exist merging vertices where any number of Pomerons disappear. Thus, for example one contribution to two Pomeron propagator would be a diagram where each one of the two Pomerons split into two, and subsequently the resulting four Pomerons merge into two via $V_2^4$. At any rate the enhanced diagrams are those where both the splitting and the merging vertices are close to each other in rapidity. The size in rapidity of a fluctuation from $n$ to $m$  Pomerons is given by the usual "time-energy" uncertainty relation $\Delta Y\sim \frac{1}{\Delta^0_m-\Delta^0_n}\sim \frac{1}{\bar\gamma}$.  "Integrating out" this time scale leads to "renormalization" of the $n\rightarrow n$ vertices, generating effective vertices $\tilde V_n^n$. In the second step we would have to resum all diagrams with insertions of these effective $\tilde V_n^n$ vertices. Those diagrams yield corrections to the $n$-Pomeron intercepts. 

The resulting values of $n$-Pomeron intercepts in principle can be either smaller or larger than their bare values. The latter option is problematic in view of the fact that in order to calculate the $S$ matrix one has to sum over all dressed Pomeron exchanges. If the $n$-Pomeron intercept grows faster than $n$, such a series is not Borel summable \cite{BORSUM}, and making sense of such an expansion presents a serious challenge. In fact one could be sufficiently discouraged to  believe that the model itself is not well defined.
In QCD, for example the simple $1/N_c^2$ corrections ($ N_c$   is the number of colors)  lead to the intercepts which are larger than the intercepts of n-BFKL Pomerons\footnote{ The very brief review of the problem, references and the most pessimistic point of view can be found in Ref.\cite{LENC} .} . Many experts therefore believe that the Pomeron calculus in QCD has a deadly problem in the order $1/N^2_c$  (including  one of us \cite{LENC}). 

In this context UTM provides an argument against such pessimism. Although we are not going to perform explicitly perturbative calculation of the type described above in UTM, from the results presented later in this section it will become clear that the perturbative Pomeron calculation does lead to positive corrections to the $n$-Pomeron intercepts. In fact, as we will show these intercepts behave as $\Delta_n=e^{\gamma n}-1$ and grow exponentially at large $n$. Nevertheless as we have stressed, this does not signify a problem with the model, as the $S$-matrix in UTM is well defined. We will show below, that even though standard Borel transform is not applicable to summation of the series of multi Pomeron exchanges, this series can be indeed summed with a finite and well defined result. We hope this may teach us something about dealing with a similar apparent problem of the  QCD Pomeron calculus.




     \begin{figure}[ht]
    \centering
  \leavevmode
      \includegraphics[width=16cm]{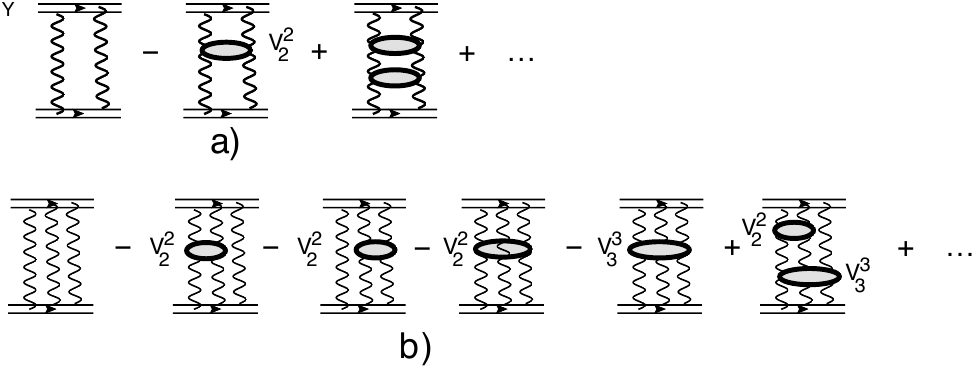}  
      \caption{ The typical Pomeron diagrams in the UTM. 
       \fig{dia}-a: the diagrams for two Pomeron Green's function. \fig{dia}-b:  the diagrams for the three Pomeron Green's function.  $\tilde V^2_2 $ is effective vertex for $\pom^2 \to  \pom^2$ interaction, while $\tilde V^3_3 $  is the effective vertex for $\pom^3 \to  \pom^3$ interaction .
       }
\label{dia}
   \end{figure}
  In the rest of this section we discuss the solution of the "Shroedinger equation" \eqref{ZEQG} and the calculation of the physical scattering matrix.
  
   \subsection{The Eigenfunctions}
   We now proceed to solve UTM along the same lines as the BFKL cascade model in the previous section. The first step in this approach is to find the eigenfunctions of the UTM Hamiltonian.
    \subsubsection{Right eigenfunctions}

It has distinct sets of right and left eigenfunctions. The right eigenfunctions are relevant for solving the equation for the generating function, or $S$-matrix. The left eigenfunctions are also interesting in their own right, since as we will show below, they are relevant for expanding the probabilities.
 
 We consider the right eigenfunctions first.
Let   $\Phi_n\Lb u,\gamma\Rb$ denote the set of eigenfunctions with positive eigenvalues $\Delta_n$:
 \beq\label{ZEQG0}
{\cal H}_{UTM}\,\Phi_n(u)\,=\,\Delta_n\,\Phi_n(u)\,~~\rightarrow\,~~ \Delta_n\,\, \Phi_n\Lb u\Rb\,\,=\,( u - 1) \Bigg(\Phi_n\Lb u\Rb\,\,- \,\Phi_n\Lb e^{-\gamma}\,u \Rb\Bigg);\eeq 
By inspection we find a set of positive eigenvalues  $\Delta_n = e^{\gamma\,n}\,-\,1$ with the eigenfunctions
\beq \label{ZEQG2}
 \Phi_{n\ge 1}\Lb u\Rb\,\,=\,\,\prod^{n-1}_{l=0} \Lb e^{ - \gamma\,l} \,\,-\,\,\frac{1}{u}\Rb\,,~~~~~~~~\Phi_0 \Lb u\Rb\equiv1
 \,,~~~~~~~~~~{\rm and }~~~~~~~~~~ \Delta_n = e^{\gamma\,n}\,-\,1\,.
 \eeq
 Notice that $\Phi_{n}\Lb u\Rb$ are well behaved near $u=1$. In particular
 $\Phi_{n>1}\Lb u=1\Rb=0$. On the other hand, $\Phi_{n}\Lb u\Rb$ diverge at $u\rightarrow 0$. 
 The functions $\Phi_n$ share these properties with $\Phi_n^{BFKL}$. However as opposed to $\Phi_n^{BFKL}$ which has a multiple pole at $u=0$ for $n>1$, $\Phi_n$ has a single pole for any $n$.
 The functions 
 $\Phi_{n}\Lb u\Rb$ could be expressed in terms of  the q-Pochhammer symbols (see formula {\bf 6.1.22} of Ref.\cite{AS} and Ref.\cite{QP}).
 
 In the formal limit $\gamma\rightarrow 0$ one has $\Phi_n(u)\rightarrow \Phi_n^{BFKL}(u)$, and also $\Delta_n\rightarrow\Delta^0_n$. Since $\Delta_n$ is perturbatively expandable in $\gamma$, this is the multi-Pomeron intercept one would obtain from perturbative approach described in the previous subsection. We thus indeed see that the resummed perturbation theory leads to n-Pomeron intercept that grows with $n$ exponentially, as advertised in the previous subsection. Our goal in calculating the $S$-matrix is then to sum the series in $n$-Pomeron exchanges. 
 
 Formally representation of the $S$-matrix in terms of these exchanges is equivalent to expansion of the generating function that solves \eqref{ZEQG} in the basis of eigenfunctions $\Phi_n$:
\beq \label{ZEQG1}
Z\Lb u,  \Y \Rb\,\,=\,\,\sum^\infty_{n=0} C_n\,\,\Phi_n\Lb u\Rb e^{ \Delta_n\,\Y}  \,. \eeq
In this expansion the boundary conditions at $u=1$ are trivially satisfied by choosing $C_0=1$. 

To find the coefficients in the expansion eq.\eqref{ZEQG1} we need to find the set of left eigenfunctions $\tilde \Psi_n$  and (with proper normalization) use
 \beq\label{sprod}
 \langle \tilde \Psi_n|\Phi_m\rangle\,=\,\delta_{mn}
 \eeq
 with appropriate definition of scalar product.
 The result, that we derive below, is that for a single dipole projectile
 \beq
C_n\rightarrow C_n^{(1)}=e^{\frac{\gamma}{2}n(n+1)}.
\eeq
We can understand a little better the nature of the series in eq.\eqref{ZEQG1} by scrutinizing the small $\gamma$ limit, and considering the physical point $u=e^{-\gamma}$ (which corresponds to the single dipole target). In this limit we have
\begin{equation}
\Phi_n(e^{-\gamma})\approx (-1)^n\gamma^nn!; \ \ ~~~~~~\ \ n\ll 1/\gamma
\eeq
and
\beq\label{A1}
\Phi_n(e^{-\gamma})\approx (-1)^ne^{\gamma\left(n-\frac{1}{\gamma}\right)}; \ \ ~~~~~\ \ \ \ \ n\gg 1/\gamma
\eeq
The $n$-th term in the expansion \eqref{ZEQG1} is $(-1)^nC_n\Phi_ne^{\Delta_n\tilde Y}$. As opposed to the BFKL case, where the analogous series was convergent for small values of $\tilde Y$, in the UTM this is an asymptotic series even for vanishing $\tilde Y$. The magnitudes of the successive terms decrease until about $n\sim1/\gamma$, and thereafter increase very fast due to the Gaussian growth of $C_n$. This behavior persists at small $\tilde Y$ until  the rapidity parametrically reaches the critical value $\tilde Y_c\approx\frac{1}{\gamma}\ln \frac{1}{\gamma}$. For larger rapidities the terms in the series grow starting at $n=1$ and the series is divergent. 
So quantitatively this is similar to BFKL in the sense that at small $Y$ the series can be truncated, although this time it is not convergent but asymptotic. At large $\tilde Y$ the series is very problematic, as the eigenvalues $\Delta_n$ grow exponentially, and thus a clever resummation is necessary to make sense of the result.

Just like in the BFKL case, the operator ${\cal H}_{UTM}$ also has another set of eigenfunctions, which we denote as $\Psi_n\Lb u\Rb$, with negative eigenvalues $-\tilde \Delta_n$.
\beq\label{ZEQG01}
{\cal H}_{UTM}\,\Psi_n(u)\,=\,-\,\tilde \Delta_n\,\Psi_n(u)\,~~~\rightarrow\,~~~ -\tilde \Delta_n\,\, \Psi_n\Lb u\Rb\,\,=\,( u - 1) \Bigg(\Psi_n\Lb u\Rb\,\,- \,\Psi_n\Lb e^{-\gamma}\,u\Rb\Bigg)
\eeq 
The eigenvectors corresponding to the negative  eigenvalues are
\beq \label{ZEQG21}
 \Psi_{n\ge 1}\Lb u\Rb\,\,=\,\prod^{n}_{l=1} \Lb e^{  \gamma\,l} \,\,-\,\,{1\over u}\Rb^{-1}\,,
 ~~~~~~~~~~{\rm and }~~~~~~~~~~ \tilde \Delta_n\, =\, 1\,-\,e^{-\gamma\,n}\,.
 \eeq
 The functions $\Psi_n(u)$ are well behaved both at $u=0$ and $u=1$, however they have poles at the physical points $u=e^{-\gamma k}$: in fact  $\Psi_n(u)$ has $n$ poles at $u_k=e^{-\gamma\,k}$ for all integer $k$ in the interval $(1\le k \le n)$.
This is significant as we will see in a short while.

The most interesting point for us that corresponds to a single dipole scattering is $u=e^{-\gamma}$.  At this point all the functions are singular with the behavior at the pole
 \beq
\Psi_n(u\rightarrow e^{-\gamma})=\frac{1}{u-e^{-\gamma}}\ e^{-\frac{\gamma}{2}(n+2)(n-1)}\prod_{l=2}^{n}\frac{1}{1-e^{-\gamma(l-1)}}
\eeq
Decomposing the general solution of  \eq{ZEQG}
 in this basis, we write
 \beq \label{ZEQG11}
Z\Lb u,  \Y \Rb\,\,=\,\,\sum^\infty_{n=1} B_n\,\,\Psi_n\Lb u\Rb e^{- \tilde\Delta_n\,\Y}  \,. \eeq

We note that the fact that $\Psi_n(u)$ have poles at physical values of $u$ does not by itself mean that the expansion of the generating function in this basis is nonsensical. As long as all the poles are of the same order (and in our case all are indeed simple poles) a simple rescaling by a common (divergent) factor will make the values of $\Psi_n$ at all the physical points finite, and everywhere else vanishing. We will not pursue this strategy here and instead continue working with the functions $\Psi_n(u)$ defined above. The price we will have to pay however, is that  we will not be able to determine the coefficients $B_n$ by themselves, but only their contribution to the physical $S$-matrix.




 

 \subsubsection{Left eigenfunctions}
 
 Let us now consider left eigenfunctions of  ${\cal H}_{UTM}$.  
 These are the same as right eigenfunctions of 
 the operator ${\cal H}_{UTM}^\dagger$ (notice the anti-Hermitian property of the dilatation operator):
 \beq\label{L1}
{\cal H}^\dagger_{UTM} =-\left(1-e^{\gamma u\frac{\partial}{\partial u}}\right) (1-u)\,,
\eeq 
Explicitly the equations satisfied by the left eigenfunctions are
\beq\label{geq} 
\lambda_n g_n(u)\,=\,(u-1)\,g_n(u)\,-\,(ue^\gamma-1)\,g_n(ue^\gamma)
\eeq
We again find two sets of solutions
\begin{eqnarray} \label{ZEQG3}
  \tilde\Phi_n\Lb u\Rb\,= 
 {1\over (1-u)} \,\prod^{n-1}_{l=0} \Lb e^{  \gamma\,l} \,\,-\,\,\frac{1}{u}\Rb\,,~~~~~\tilde\Phi_0 \Lb u\Rb\,=\, (1-u)^{-1} \,,
 ~~~~~\lambda_n= -\,\tilde \Delta_n \,=\, -\,(1\,-\,e^{-\gamma\,n}).
 \end{eqnarray}
 and
\beq\label{ZEQG23}
\tilde \Psi_n\Lb u\Rb\,\,=
-\,
{1\over u}\,\prod^{n}_{l=0} \Lb e^{ - \gamma\,l} \,-\,{1\over u}\Rb^{-1}\,,
 ~~~~~~~~~~ \tilde\lambda_n\,=\, \Delta_n\, =\,e^{\gamma\,n}\,-\,1.
 \eeq
 The left eigenfunctions are regular for physical values of $u$, i.e. $0\le u\le 1$, with $\tilde\Phi_n(e^{-\gamma k})=0$  for $k<n$. On the other hand for $u>1$ the functions $\tilde\Psi_n(u)$ have poles at $u=e^{\gamma k}$ for $k\le n$.
 
 For the record we note that $\tilde \Phi_n$ obey two recurrent relations:
\beq
\tilde \Phi_k\Lb e^{-\gamma n}\Rb\,=\, \left[e^{\gamma\,(k-1)} \,-\,e^{\gamma\, n}\right] \,\tilde \Phi_{k-1}\Lb e^{-\gamma n}\Rb\,,
~~~~~~~\tilde \Phi_k\Lb e^{-\gamma n}\Rb\,=\,{1-e^{-\gamma\,(n-1)}\over e^{-\gamma\, k}- e^{-\gamma\, n}} \,\tilde \Phi_k\Lb e^{-\gamma(n-1)}\Rb\,
\eeq

\subsubsection{Expansion Coefficients $C_n$ and Exponential Moments}
To calculate the expansion coefficients $C_n$ in \eqref{ZEQG1} we use a somewhat roundabout procedure. Although the generating function $Z(u)$ in the physical region (where it represents the scattering matrix)  is defined for $0\le u\le 1$, we can consider it as a function of all real $u$. We assume that the expansion \eqref{ZEQG1} holds for all $u$ with the same coefficients $C_n$. 

Define an auxiliary variable $\xi_n\equiv e^{\gamma n}$. The value of $Z$ at the points $u=e^{\gamma k}$ is equal to the $k$-th exponential moment:
\beq \label{MOXI}
\,Z(e^{ \gamma\,k},Y)=M_k\Lb Y\Rb \,\equiv\, \eval{(\xi_n)^k}\,\,=\,\,\sum^\infty_{n=1} \xi_n^k\,P_n\Lb Y\Rb\,
\eeq
The moments are expanded in terms of the right eigenfunctions as
\beq \label{MOXI}
 M_k\Lb Y\Rb  \,=\,
 \sum^\infty_{n=0}  C_n\,\,\Phi_n\Lb \xi_k\Rb\, e^{\Delta_n\,\Y} \,=\,
  \sum^{k}_{n=0}  C_n\,\,\Phi_n\Lb \xi_k\Rb \,e^{\Delta_n\,\Y}
\eeq
where the last equality holds since $\Phi_n(\xi_k)=0$ for $n>k$ as is obvious from \eqref{ZEQG2}.
To find the coefficients $C_n$ we introduce  the inverse matrix $Q_{m}^k$ that satisfies
\beq
\sum_{k=0}^{\infty} Q_m^k\, \Phi_n\Lb \xi_k\Rb\,=\,\sum_{k=n+1}^{\infty} Q_m^k\, \Phi_n\Lb \xi_k\Rb\,=\,\delta_{nm}
\eeq
Hence
\beq
\sum_k Q_m^k\,M_k\,=\, C_m\,e^{\Delta_m\,\Y}
\eeq
We now utilize the theorem about the orthogonality of left and right eigenfunctions. It is straightforward to check that the appropriate measure under which this orthogonality holds is
\beq \label{ort3}
\frac{1}{2\pi i}\oint_{\,\Gamma_3} {du\over u}\, \tilde\Psi_m(u)\,  \Phi_n\Lb u\Rb\,=\,\delta_{mn}
\eeq
Here the contour $\Gamma_3$ encloses the real axis to the right (and including) the point $u=1$, and thus encircles all the poles of $\tilde \Psi_m$, (see Fig. \ref{contga}).
Using this relation we find
\beq
C_n\,=\frac{1}{2\pi i}\,\oint_{\,\Gamma_3} {du\over u}\, \tilde\Psi_n(u)\,  Z\Lb u,Y=0\Rb
\eeq
For the $k$-dipole target  initial condition, $Z\Lb u,Y=0\Rb\,=\,u^k$ this is:
\beq
C_n^{(k)}\,=\,\prod_{l=1}^{n} {e^{\,l\,\gamma}\,(e^{\,\gamma\,(l+k-1)} -1)\over e^{\,l\,\gamma}-1}
\eeq
For $k=1$, 
 \beq
 C_n^{(1)}\,=\,e^{\gamma\,n(n+1)/2}\,.
 \eeq 
 which is the result quoted in the beginning of this section.

\subsection{The Probabilities}

One interesting property of the operator ${\cal H}_{UTM}$ is that it's action  on a function of continuous variable, $\phi(u)$ is such that it only relates the values of the function at discrete values of its argument $\phi(e^{-\gamma k}u)$. This is very different from an action of a differential operator, such as ${\cal H}_{BFKL}$ \eqref{lbfkl}. Thus, although we have been treating the eigenfunctions as continuous functions of continuous variable, strictly speaking  the eigenvalue equation only determines eigenfunctions at discrete points $e^{-\gamma k}u_0$ for an arbitrary value of $u_0$. Since the point $u=1$ is a physical point for scattering amplitude, our calculations in the previous sections corresponded to choosing $u_0=1$ and determining the generating function at points $u_k=e^{-\gamma k}$. The value of the generating function at these particular values is indeed physically significant, since it is equal to the scattering matrix for a $k$-dipole target. The values of any eigenfunction at other points, excluding this discrete set, is to some extent arbitrary and were chosen to make functions continuous.

We can use this property of  ${\cal H}_{UTM}$ to cast the equation for probabilities in the form similar to that for the generating function. Let us formally extend the probabilities to a "probability function" - a function of continuous variable such that
\begin{equation}
P_n(\tilde Y)\rightarrow P(u,\tilde Y); \ \ \ \ ~ ~ ~~~~~~\ \ \ P(u=e^{-\gamma n},\tilde Y)=P_n(\tilde Y)
\end{equation}
The equation (\ref{MEQ}) for $P^{UTM}_n$ can be rewritten as an equation for the function $P(u)$
\beq \label{MEQ1}
\frac{d P(u,\tilde Y)}{ d \tilde Y}\,=\, {\cal H}_{UTM}^\dagger \, P(u,\tilde Y)  ~~~~~
\eeq
The function $P(u)$ can be expanded in the set of left eigenfunctions of ${\cal H}_{UTM}$,
\beq \label{PEQG11}
P\Lb u,\Y \Rb\,=\,\sum^\infty_{k=0} \tilde C_k\,\tilde \Phi_k\Lb u\Rb \,e^{-\tilde\Delta_k\,\Y} 
=\,\sum^\infty_{k=1} \tilde B_k\,\tilde \Psi_k\Lb u\Rb\, e^{\Delta_k\,\Y}  \,. \eeq

The coefficients $\tilde C_k$ are found by using the orthogonality condition
\beq \label{ort2}
\frac{1}{2\pi i}\oint_{\Gamma} {du\over u}\, \Psi_m(u)\, \tilde \Phi_n\Lb u\Rb\,=\,\delta_{mn}
\eeq
where the contour $\Gamma$ encircles the physical interval $0<u\le 1$, excluding the origin $u=0$ (see Fig. \ref{contga}). This relation can be verified directly using the explicit form of the functions $\Psi_n$ and $\tilde\Phi_n$.

For the single dipole initial condition, at initial rapidity the function $P(u)$ should be chosen such that its value is equal to unity at $u=e^{-\gamma}$, and the function vanishes quickly away from this point along the real axis. For example $P(u,\tilde Y=0)=e^{-\left[\Lambda(u-e^{-\gamma})^2\right]}$ with $\Lambda\gg \frac{1}{\gamma}$ will do.
For such functions, in the integral $\frac{1}{2\pi i}\oint_{\Gamma} {du\over u}\, \Psi_k(u)P(u,\tilde Y=0)$ only the pole of $\Psi_k(u)$ 
at $u=e^{-\gamma}$ contributes, and we find
\begin{eqnarray} \label{B}
\tilde C_0&=&0\nonumber\\
\tilde C_1&=&-e^{-\gamma}\nonumber\\
\tilde C_{m}\,&=&\,\bar Q_{m>1}^1\,=\,-\,2^{(1-m)}\,e^{-\gamma\,m(m+3)/4} 
\,\prod_{l=1}^{m-1}  { 1\over \sinh[\gamma\,l/2]}; ~~~~~~\ \ \ m>1;
\end{eqnarray}

Note that since $\tilde \Phi_k\Lb e^{-\gamma n}\Rb \,=\,0$ for all $k>n>0$, the probability function at physical points where it is equal to the dipole probability in the cascade, is in fact a finite sum
\beq \label{PEQG12}
P_n^{UTM}\Lb \Y \Rb\,=\,\sum^n_{k=1} \tilde C_k\,\tilde \Phi_k\Lb e^{-\gamma n}\Rb \,e^{-\tilde\Delta_k\,\Y} 
\eeq

In principle we can find similarly the coefficients $\tilde B$ in the second expansion in \eqref{PEQG11}. However it is not clear to us that expansion of the probability function  in positive exponents of rapidity  can ever be useful, and we will not do it here.

  \section{Dipole  scattering in the UTM}
 
 Our goal now is to understand the behavior of the $S$-matrix in the UTM and to show that notwithstanding apparent problems with Pomeron calculus it can be calculated. In particular we will explicitly calculate the asymptotic behavior at large $\tilde Y$.
 
 Since UTM is   frame independent by construction, below we carry calculations in the frame which is the 
 most convenient for us, and that is the target rest frame.

  \subsection{Scattering on a target dipole}
  
  From \eq{SMS}, the $S$-matrix on a single target dipole reads
 \beq \label{SSS}
S\Lb \Y\Rb\,\,=\,\,Z\Lb e^{- \gamma},\Y\Rb \,\,=\,\,
\sum^\infty_{n=1} e^{-\gamma\,n}\,P_n^{UTM}\Lb \tilde Y\Rb 
\eeq
 Using \eqref{PEQG12} and expanding in powers of $\tilde Y$ we can write it as
 \beq 
\label{sp} S(\tilde Y)= e^{- \,\Y}\sum^\infty_{j=0} \sum_{n=1}^{j+1}
\frac{\Y^j}{j!}
e^{-\gamma} \,{e^{-\gamma n j}} \prod^{n-2}_{l=0} \Lb1\,\,-\,\,e^{\gamma n(l-j)} \Rb\
\eeq
We will use this expression for numerical evaluation in the following.
 
 In the previous section we have provided two expansions for $S(\tilde Y)$ - expansion in multi Pomeron amplitudes and expansion in decreasing exponents of rapidity. We will
 show in this section that both lead to the same asymptotic result for the $S$-matrix. We start with the expansion in Pomeron exchanges.
 
 \subsubsection{Pomeron calculus}
We start with the representation of $S$-matrix in terms of "dressed" multi-Pomeron exchanges 
\beq \label{SA}
S\Lb \Y\Rb\,\,=\,\,\sum^\infty_{n=0} C_n^{(1)}\,\Phi_n\Lb e^{- \gamma}\Rb\, e^{ \Delta_n\,\Y}\
\eeq
Here $\Delta_n=e^{\gamma n}-1$ is the $n$-Pomeron intercept (in $\Y$ variable)
\footnote{Naively one might want to identify  $C_n\Phi_n$ as the product of the residues (impact factors), with  $C_n$ being related to  the projectile while $\Phi_n$ to  the target. However, residues factorize for bare single Pomeron exchange only. Beyond one Pomeron exchange the factorizations breaks down. }.

 \eq{SA} is an asymptotic series, which, at first glance, cannot be summed because the coefficients grow too fast with $n$:
$\Delta_n \,>\,\gamma \,n$, while $C_n^{(1)}\Lb \gamma\Rb = \exp\Lb \frac{n (n+1)}{2}\,\gamma\Rb$  grows as Gaussian, which is badly divergent too \footnote{In Appendix \ref{App A} we suggest  how to sum the asymptotic series with the
Gaussian growth of the coefficients.}.
In addition $\Phi_n$ also grows with $n$ exponentially. The only redeeming feature of this expression is that the terms in the series are alternating sign since which is clear from 
\eqref{ZEQG2}.

Summing  series of the type \eq{SA} is a long standing problem in the Pomeron calculus (see Ref.\cite{LENC} for a short review). 
Summation of  asymptotic series entails finding an  analytical function with identical series expansion.  
 Borel summation \cite{BORSUM} is one of the standard tools of the trade in this business. The series \eq{SA} however cannot be summed via Borel procedure
 due to the aforementioned  very fast growth of the coefficients. We will therefore rearrange the series in the following way.
We first expand the exponential factors using the explicit form of
 $\Delta_n$ as
\beq \label{EXP}
\exp\Lb \Delta_n\,\Y\Rb\,\,=\,\, e^{- \,\Y}\sum^\infty_{j=0} \frac{e^{\gamma \,n\,j}\,\Y^j}{j!}. 
\eeq
Substituting \eq{EXP} in \eq{SA} 
we obtain:
\begin{eqnarray} \label{SA11}
S\Lb \Y\Rb\,\,=\,\,Z\Lb e^{- \gamma},\Y\Rb &=&e^{- \,\Y}\sum^\infty_{j=0} \sum^\infty_{n=0}\frac{e^{\gamma \,n\,j}\,\Y^j}{j!}C_n^{(1)} \Phi_n\Lb e^{-\gamma}\Rb  
\end{eqnarray}
In \eq{SA11} we have two series: an absolutely convergent sum over $j$ and 
 an asymptotic series over $n$. 
     \begin{figure}[ht]
    \centering
  \leavevmode
  \begin{tabular}{c c c}
      \includegraphics[width=7.9cm]{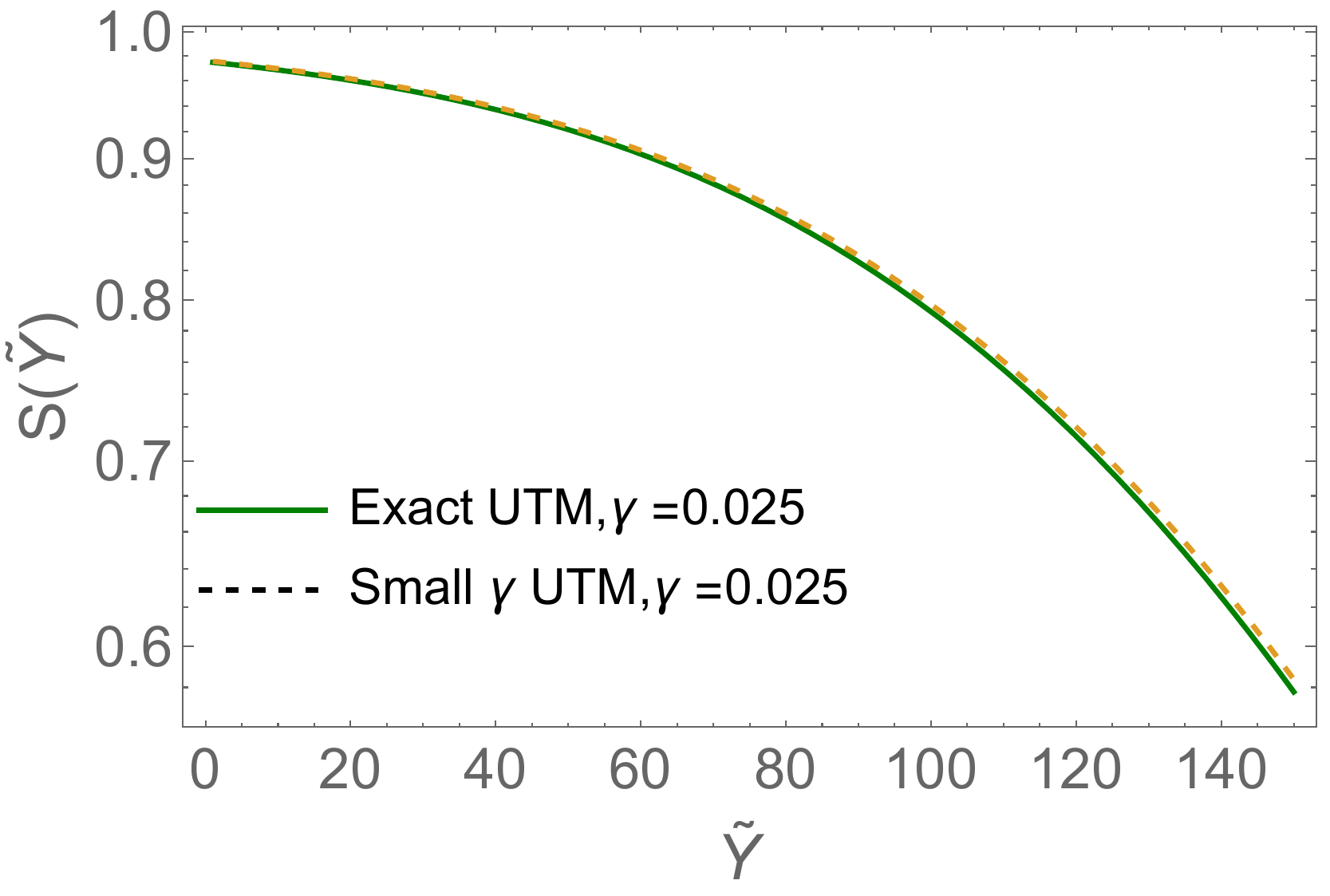}& ~~~~&  \includegraphics[width=8.1cm]{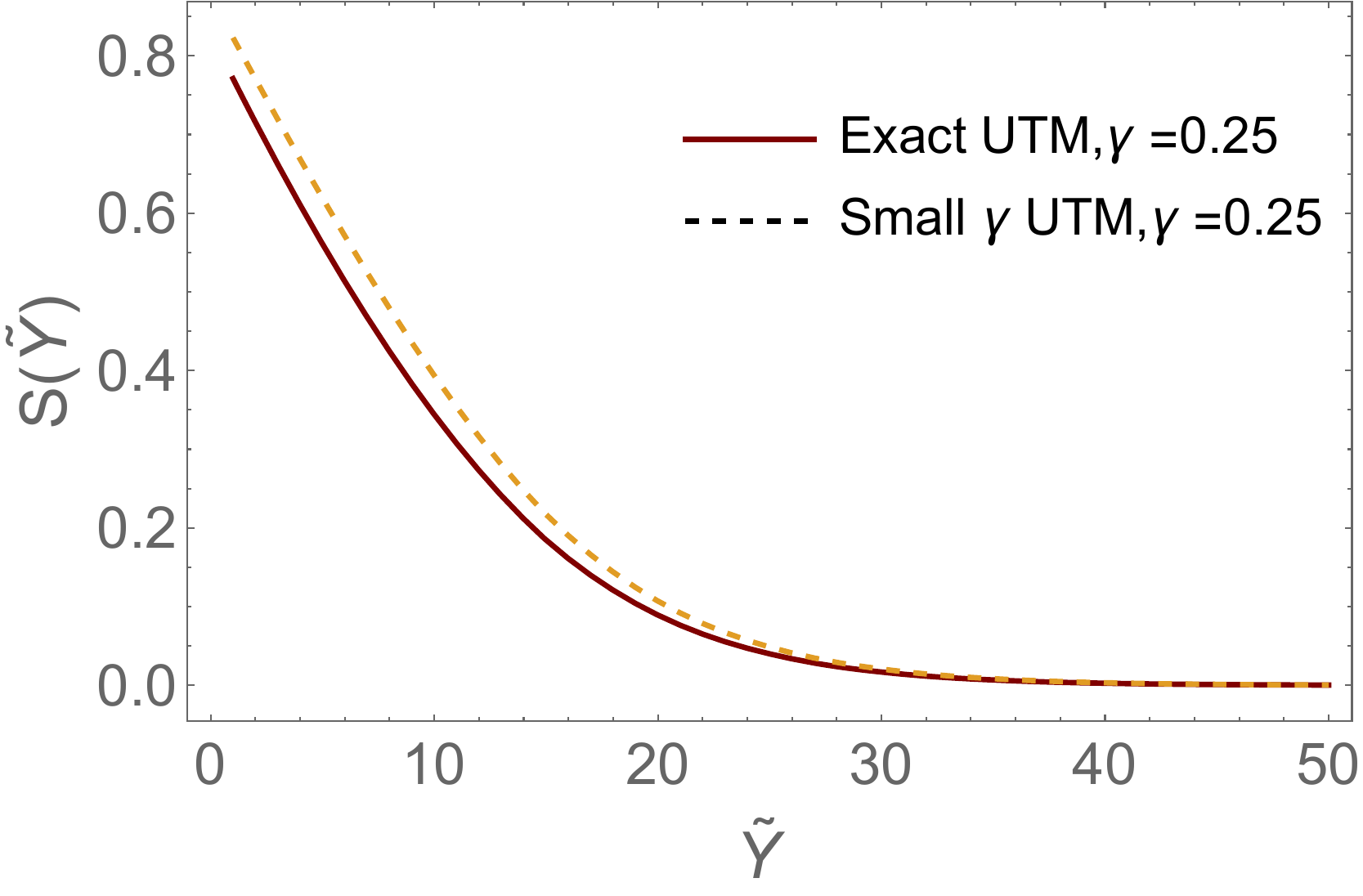} \\
      \end{tabular}
           \caption{ $S\Lb \Y\Rb$ versus $\Y$ for  (i) exact solution of \eq{sp}, and (ii)   small $\gamma$ approximation  of \eq{SA5}.  For $\gamma=0.025$ (left plot), the  curves coincide with $0.1\%-1\%$ accuracy. For $\gamma=0.25$ (right plot), 
             the accuracy ranges from $5\%$ to $20\%$ in the displayed interval, worsening at larger rapidities.
      }
\label{ampl}
   \end{figure}

\subsubsection*{Small $\gamma$ limit}

We first explore the limit  $\gamma\ll 1$. For small $\gamma$ we have  $C_n^{(1)}\simeq 1$ and 
$\Phi_n\Lb 1 - \gamma\Rb \,\,=\,\,\Lb - \gamma\Rb^n \,\,n!\,\,+\,\,O\Lb \gamma^{n+1}\Rb$. Keeping only leading terms we can write
\beq \label{SA2}
S\Lb \Y\Rb\,\,=\,\,
\,\,\,\,\sum^\infty_{n=0} \Lb - \gamma\Rb^n \,\,n! \,\,e^{ \Delta_n\,\Y}\,\,\Bigg\{ 1\,\,\,+\,\,{\cal O}\Lb \gamma\Rb\Bigg\}\,,\,\quad\quad\, 
\gamma\ll1\,.
\eeq
Neglecting subleading terms may seem risky in view of our discussion in Section 2. In particular keeping only the leading term for $C_n$ we omit the Gaussian enhancement at large $n$. However it is the presence of exponentially growing $\Delta_n$ that poses the main problem for resummation, and we keep $\Delta_n$ in full glory in \eqref{SA2}.

If we were to expand $\Delta_n$ for formally small $\gamma$, that is $\Delta_n\simeq n\gamma$, we would reproduce  the large $Y$ limit of the BFKL cascade model in the center of mass frame
(\ref{SMS41d}).

Applying the expansion (\ref{EXP}) to \eqref{SA2} we have,
\beq
S\Lb \Y\Rb\,
\,=\,\,\,e^{- \,\Y}\sum^\infty_{j=0} \sum^\infty_{n=0}\frac{e^{\gamma \,n\,j}\,\Y^j}{j!}\,\Lb - \gamma\Rb^n \,\,n!
 \eeq
The asymptotic series in $n$ is Borel summable and gives
\beq \label{SA5}
S\Lb \Y\Rb\,\,=\,\,
\,\,e^{- \,\Y}\sum^\infty_{j=0} \frac{\Y^j}{j!}
\frac{1}{\gamma}e^{-(j+1)\gamma} \exp\Lb   \frac{1}{\gamma}e^{-(j+1)\gamma}  \Rb \,\Gamma\Lb 0,\frac{1}{\gamma}e^{-(j+1)\gamma} \Rb
\eeq
For  large $\Y \gg 1$, the sum is dominated by terms with large $j$, and    $ \frac{1}{\gamma}e^{-(j+1)\gamma}\ll 1$. Taking the asymptotic form of the incomplete $\Gamma$-function at small argument, we get
\bea \label{SA51}
S\Lb \Y\Rb\,\,&=&\,\,
\,\,e^{- \,\Y}\sum^\infty_{j=0} \frac{\Y^j}{j!}\Lb \frac{1}{\gamma \xi_j}\Lb  \ln\Lb \gamma \xi_j\Rb\  \,-\gamma_E - \frac{1}{4}\gamma\Rb\Rb \nn\\
&\,=\,&e^{ - ( 1 - e^{-\gamma}) \Y}\Lb e^{- 2 \gamma} \,\Y + e^{ - \gamma} \Lb \frac{1}{\gamma} \Lb \ln \gamma - \gamma_E\Rb +\frac{3}{4} \Rb\Rb
\eea
  $\gamma_E=0.577$ is the Euler constant.  Eq. (\ref{SA51}) gives the leading asymptotic behavior of the $S$-matrix at 
 small $\gamma$. The validity of this approximation is demonstrated in \fig{ampl} alongside the exact result. We observe 
 that even for $\gamma=0.25$ the small $\gamma$ approximation is very good unless  extremely large rapidities are considered.
 
 Note that even though we started with expansion of $S$ in multi-Pomeron exchanges which grow exponentially, the final result \eqref{SA51} is given by the first decreasing exponent $-\tilde\Delta_1$. This suggests that, as in the BFKL limit at large rapidity the appropriate expansion should be that in the negative exponents $e^{-\tilde\Delta_n\tilde Y}$ rather than in the multi-Pomeron exchanges.
 However, the presence of the prefactor $\tilde Y$ in \eqref{SA51} indicates that a straightforward expansion of $S$-matrix  in this basis is not possible, as the asymptotics is not given by a pure exponential.  Indeed let us try to find coefficients $B_n$ in \eqref{ZEQG11}. For this we should use the following orthogonality condition
  \beq\frac{1}{2\pi i}\oint_{\Gamma}\frac{1}{u}\Psi_m(u)\tilde \Phi_m(u)=\delta_{mn}
  \eeq
However, naively calculating the overlap with the initial condition $Z(u,\tilde Y=0)=u$ we find
  \beq
  B_m=\frac{1}{2\pi i}\oint_{\Gamma}\frac{1}{u}u\ \tilde \Phi_m(u)=0
  \eeq
  as the integrand has no poles to the right of $u=0$. 
This result is consistent with our observation below \eqref{ZEQG11}, but is not helpful for finding the $S$-matrix.
Thus to find  a better approximation for the $S$-matrix we need to look for a somewhat more refined  method than just formal expansion in the basis $\tilde\Psi(u)$. In the rest of this section we develop such a method based on the analytic continuation.

\subsubsection{The analytic continuation}

We start with the expansion of the generating function in multi-Pomeron exchanges, but represent it formally as an integral in the complex $n$-plane:
\beq\label{zan}
Z\Lb u,  \Y \Rb\,\,=\,{1\over 2\pi i}\,\oint_{C} {(-1)^n \pi dn\over \sin(\pi \,n)}\, C_n^{(1)}\,\,\Phi_n\Lb u\Rb e^{ \Delta_n\,\Y} 
\eeq
where the contour $C$ runs along the real positive semi-axes (see \fig{contr}). 
Let us assume that all the functions of $n$ can be analytically continued to the complex plane in such a way that the integrand vanishes fast enough at infinity.  In this case there is no contribution from the circle at infinity and 
we can deform the integration contour and put it along the real negative semi-axes, the contour $C^{\prime}$.
     \begin{figure}[ht]
    \centering
  \leavevmode
      \includegraphics[width=6cm]{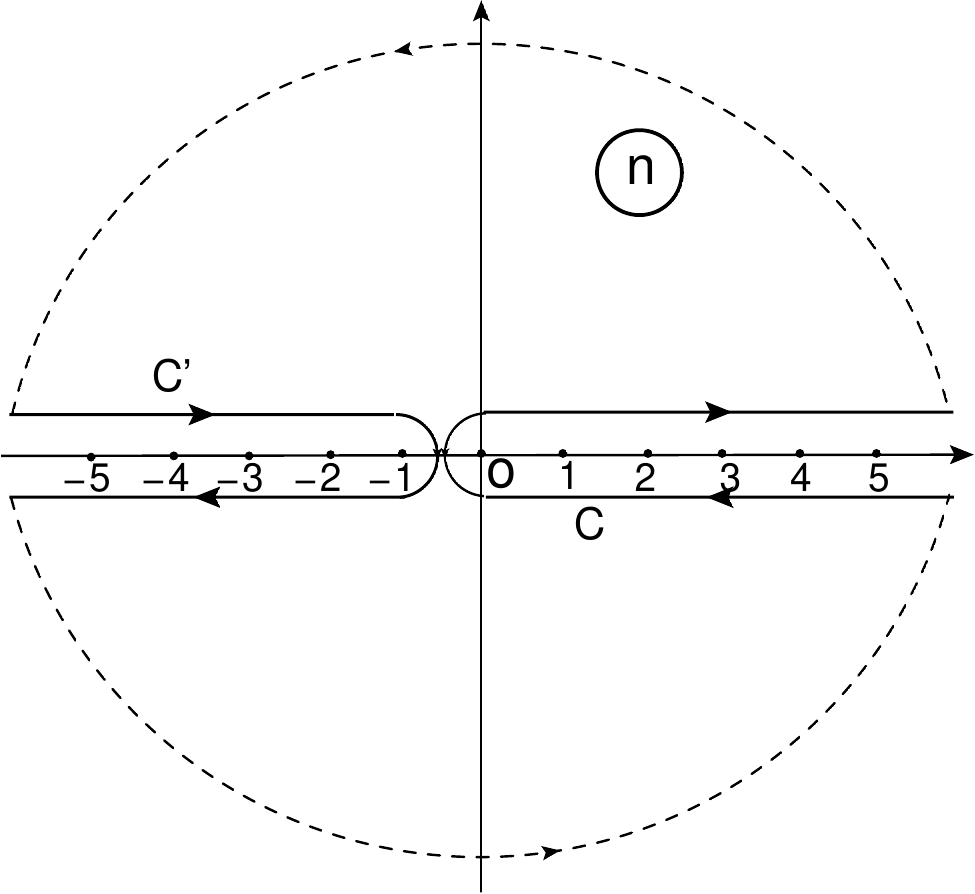}  
      \caption{ Contours of integration in the complex $n$ plane.}
\label{contr}
   \end{figure}
%

To integrate along $C'$ we only need to know the analytic continuation of the integrand in \eqref{zan} close to the negative real axis. To find this analytic continuation 
 we first note that 
eigenfunctions $\Phi_n$ satisfy the recurrence relation
\beq
 \Phi_{n+1}\Lb u\Rb\,=\, \Phi_{n}\Lb u\Rb\,\left(e^{-\gamma\,n} -{1\over u}\right)
\eeq
We can use this relation to continue the function $\Phi$ to negative integer $n$:
\beq
 \Phi_{-n+1}\Lb u\Rb\,=\, \Phi_{-n}\Lb u\Rb\,\left(e^{\gamma\,n} -{1\over u}\right)
\eeq
This can be iterated to give
\beq\label{recurrent}
 \Phi_{-n}\Lb u\Rb\,=\, \Phi_{-(n-1)}\Lb u\Rb\,\left(e^{\gamma\,n} -{1\over u}\right)^{-1}\,=\,\,\prod^{n}_{l=1} \Lb e^{  \gamma\,l} \,\,-\,\,{1\over u}\Rb^{-1}\,=\,\Psi_{n}\Lb u\Rb
\eeq
We have already noticed that  $\Psi_n(u)$ have simple poles in $n$ at $u=e^{-\gamma\,k}$, for all integer $k\le n$. 
Thus in the vicinity of integer $n$ we can write
\beq\label{recurrent1}
 \Phi_{-n}\Lb u=e^{-\gamma\,k}\Rb\,=\, \Psi_{n}\Lb u=e^{-\gamma k}\Rb\,=\,
 -\,{(-1)^n \pi\,\tilde Q_{n}^k\over \gamma\,\sin(\pi\,n)}\,
\eeq
where $\tilde  Q_{n}^k$  are  introduced as  residues of the poles in $n$. Our goal now is to find these residues.

For a single dipole target we need  $k=1$:
\bea \label{SA500}
 \Phi_n\Lb e^{-\gamma}\Rb &= &(-1)^n\,e^{\gamma\,n} \prod^n_{l=1}\Lb 1 \,-\,e^{ -\gamma\,l}\Rb =(-1)^n\,e^{\gamma\,n -  \gamma\frac{n (n+1)}{4}}  \prod^n_{l=1}\Lb 2 \sinh\Lb \h \gamma l\Rb\Rb 
\\
 &=&(-\gamma)^n\,n!\,e^{\gamma\,n -  \gamma\frac{n (n+1)}{4}}\exp\Lb \sum^n_{l=1} \ln\Lb \frac{2 \sinh\Lb \h \gamma l\Rb}{ \gamma l}\Rb\Rb\nn\\
 &=&
 e^{\gamma\,n -  \gamma\frac{n (n+1)}{4} }( - \gamma )^n n!  \exp\Lb  \sum^\infty_{k=1}\sum^n_{l=1}  \mu_k( \gamma\,l)^{2 k}\Rb=
 e^{\gamma\,n -  \gamma\frac{n (n+1)}{4} }( - \gamma )^n n!  \exp\Lb  \sum^\infty_{k=1} \mu_k\, \gamma^{2 k } H_n^{(-2 k)}\Rb 
  \nn 
  \eea 
  Here $H_n^k$ is the generalized harmonic number.
 Coefficient $\mu_k$ can be easily calculated (see Ref.\cite{RY}, {\bf 1.518(1)}):
 \bea \label{SA5001} 
 \mu_k \,=\, \frac{2^{k+1}\,{\cal  B}_{2k}}{2 k\,(2 k)! }\,=\,-\, (-1)^k\, \frac{\zeta\Lb 2 k\Rb}{k\,(2 \pi)^{2k} }
 \eea
 where ${\cal B}_{2k}$  are Bernoulli numbers.
Thus we get
 \beq \label{SA5002}
 \Phi_n\Lb e^{-\gamma}\Rb = e^{\gamma\,n -  \gamma\frac{n (n+1)}{4} }( - \gamma )^n \Gamma\Lb n+1\Rb  \exp\Lb  \sum^\infty_{k=1} (-1)^{k +1}\frac{1}{k}\zeta\Lb 2 k\Rb \Lb \frac{\gamma}{2 \,\pi}\Rb^{2 k } H_n^{(-2 k)}\Rb 
 \eeq
 and
 \beq \label{SA50021}
\Psi_n\Lb e^{-\gamma}\Rb\,=\, \Phi_{-n}\Lb e^{-\gamma}\Rb 
= e^{-\gamma\,n -  \gamma\frac{n (n-1)}{4} }( - \gamma )^{-n} \Gamma\Lb -n+1\Rb  \exp\Lb  
\sum^\infty_{k=1} (-1)^{k +1}\frac{1}{k}\zeta\Lb 2 k\Rb \Lb \frac{\gamma}{2 \,\pi}\Rb^{2 k } H_{-n}^{(-2 k)}\Rb 
 \eeq
The function  $\Gamma\Lb -n+1\Rb$ has poles at integer values of $n$.
Using
 $   -\,H^{(-2 k)}_{-n-1} = H_n^{(-2 k)}$ and $\Gamma(1-z)\,\Gamma(z)\,=\,{\pi\over \sin(\pi\,z)}\,
$ we can finally write
 \bea \label{SA5003} 
 \tilde Q_{n}^1\,=\,-\,
{1\over \gamma^{n-1}\,(n-1)!}\, e^{ \gamma \frac{n (n -1)}{4} - n \gamma}\exp\Lb  \sum_{k=1}^\infty
   \gamma^{2 k}\,\mu_k \,H^{(-2 k)}_{n-1}\Rb\, .
 \eea 
This expression with $(n-1)!\rightarrow\Gamma(n)$ provides the analytic continuation of $\tilde Q^1_n$ to the complex $n$ plane.
Notice that $\tilde Q^1_n\,=\,\bar Q^1_n$  computed in (\ref{B}).


The $S$-matrix then is
\begin{eqnarray} \label{MISHA3}
S\Lb \Y \Rb\,&=&{1\over 2\pi i}\
\oint_{C} {\pi\,(-1)^ndn\over \sin(\pi \,n)}\, C_n^{(1)}\,\Phi_n\Lb u=e^{-\gamma} \Rb e^{ \Delta_n\,\Y} 
\,=\,{1\over 2\pi i}\oint_{C^\prime} {\pi^2  dn\over \gamma\,\sin^2(\pi \,n)}\, C_{-n}^{(1)}\,\tilde Q_{n}^1 \, e^{ -  \tilde{\Delta}_n\,\Y}
\nonumber \\
&=&\,{1\over\gamma}\,
\sum^\infty_{n=1} {d\over dn} \left[C_{-n}^{(1)}\,\tilde Q_{n}^1 \,e^{ -\tilde\Delta_n\,\Y} \right]\,=\,
\Y\sum_{n=1} \bar\beta_n\, e^{ -\tilde\Delta_n\,\Y}\,+\,\sum_{n=1} \beta_n\, e^{ -\tilde\Delta_n\,\Y}
\end{eqnarray}
with
\beq\label{MISHA2}
\beta_n\,\equiv\,{1\over\gamma}{d\over dn} \left[C_{-n}^{(1)}\,\tilde Q_{n}^1\right]\,;~~~~~~~~~~~~~
\bar\beta_n\,\equiv\,-\,  C_{-n}^{(1)}\,\tilde Q_{n}^1\,e^{-\gamma\,n}
\eeq
In the limit $n\rightarrow\infty$,
\beq\label{ass}
\lim_{n\rightarrow \infty} \left[C_{-n}^{(1)}\,\tilde Q_{n}^1\,e^{ -\tilde\Delta_n\,\Y} \right]\,=\,-\,e^{-\gamma\,n}\,e^{ -\,\Y} 
\eeq 
thus the series in $n$ is convergent even though the coefficients $C_n$ individually have a Gaussian growth.

For large enough $\Y$ the dominant contribution comes from $n=1$. Keeping also terms that are large for small $\gamma$ we have
\beq
{1\over\gamma}{d\over dn} \left[C_{-n}^{(1)}\,\tilde Q_{n}^1\,e^{ -\tilde\Delta_n\,\Y} \right]_{n=1}\,\simeq\,-\,
{d\over dn} \left[e^{-\gamma}\gamma^{-n}\,e^{ -\tilde\Delta_n\,\Y}\right]_{n=1}\,=\,e^{-2\gamma}\,\left[ \Y \,+\,e^{\gamma}\,{1\over\gamma}\,\ln\gamma       \right]\,e^{ -\tilde\Delta_1\,\Y}
\eeq
which provides both the asymptotic and pre-asymptotic behavior consistent with Eq. (\ref{SA51}).
At finite $\gamma$, the leading large $Y$ asymptotics is
\beq
S\Lb  \Y \Rb\,=\,
{1\over\gamma} \,C_{-1}^{(1)}\,\tilde Q_{1}^1 \,{d\over dn}\,\left[e^{ -\tilde\Delta_n\,\Y}\right]_{n=1}\,=\,
e^{-2\gamma}\,\Y\,e^{ -\tilde\Delta_1\,\Y}
\eeq

After exploring the asymptotics, we  will now   keep all the subleading terms. For convenience, let us define new variable $\alpha_n$:
\beq \label{MISHA1}
{1\over\gamma}\,C_{-n}^{(1)}\Lb \gamma\Rb\,\tilde Q_{n}^1\,=\,-\,{e^{\gamma\,n^2/4 \,-\, 5\gamma\,n/4}\over \ga^{n}}\,\alpha_n\,;
~~~~~~~~~\alpha_n\,\equiv\,\prod_{l=1}^{n-1} {\gamma\over 2\sinh[\gamma\,l/2]}
\eeq
These equations define $\alpha_n$ at integer values of $n$, but we understand  it in the sense of analytic continuation.
Hence
\begin{eqnarray}\label{beta}
\bar\beta_n &=&   {e^{\gamma\,n^2/4 \,-\, 9\,\gamma\,n/4}\over \ga^{n-1}}\,\alpha_n\, \nonumber \\
\beta_n&=&    {e^{\gamma\,n^2/4 \,-\, 5\,\gamma\,n/4}\over \ga^{n}}\,\left[ {\ln(\gamma)}\, \alpha_n\,
- (n/2\,-\,5/4) \,\gamma\,\alpha_n -\alpha^\prime_n\right]\end{eqnarray}
Formally \eq{MISHA3} is an infinite series. However as is clear from \eqref{ass} the series is convergent at any value of rapidity $\tilde Y$. 
Let us introduce a separation between "large" and "small" terms in the expansion, $n_0$.
At the initial rapidity $Y=0$,
\beq
S\Lb   \Y=0 \Rb\,=\,e^{-\gamma}\,=\,\sum_{n=1}^{n_0} \beta_n\,+\,\sum_{n=n_0+1}^\infty \beta_n\,
\eeq
Hence
\beq
\sum_{n=n_0+1}^\infty \beta_n\,=\,e^{-\gamma}\,-\,\sum_{n=1}^{n_0} \beta_n
\eeq
We choose $n_0(\tilde Y)$ such that for $n>n_0$, 
\beq
e^{-\tilde\Delta_n\,\tilde Y}\, \simeq \,e^{-\tilde Y}\,(1\,{+} \,e^{-\gamma\,n}\,\tilde Y)\,;\,~~~~~~~~\,\tilde Y\,\ll\,e^{\gamma\,n_0}.
\eeq
Using this relation, we can rewrite the $S$-matrix at arbitrary rapidity $\Y$
\begin{eqnarray}
S\Lb  \Y \Rb&\simeq& e^{-\gamma}\,e^{-\tilde Y}+
\sum_{n=1}^{n_0} \beta_n\, \left[e^{ -\tilde\Delta_n\,\Y}\,-\,e^{-\tilde Y}\right]+\tilde Y\sum_{n=1}^{n_0} \bar\beta_n\, e^{ -\tilde\Delta_n\,\Y}
+\tilde Y\,e^{ -\Y}\sum_{n=n_0+1}^{\infty} [\bar\beta_n\, -\,\beta_n\,e^{-\gamma\,n}]\nonumber \\
&\simeq& e^{-\gamma}\,e^{-\tilde Y}+
\sum_{n=1}^{n_0} \beta_n\, \left[e^{ -\tilde\Delta_n\,\Y}\,-\,e^{-\tilde Y}\right]+\tilde Y\sum_{n=1}^{n_0} \bar\beta_n\, e^{ -\tilde\Delta_n\,\Y}
\end{eqnarray}
The last term in the first line can be dropped since for very large $n$, 
$ \bar\beta_n\, =\,\beta_n\,e^{-\gamma\,n}$ as follows from the 
asymptotics 
(\ref{ass}).
The very same result could be arrived at in a somewhat shorter way:
\begin{eqnarray}
S\Lb  \Y \Rb\,&=&
\Y\sum_{n=1}^{n_0} \bar\beta_n\, e^{ -\tilde\Delta_n\,\Y}\,+\,\sum_{n=1}^{n_0} \beta_n\, e^{ -\tilde\Delta_n\,\Y} \,+\,{1\over\gamma}\,
\sum^\infty_{n=n_0+1} {d\over dn} \left[C_{-n}\Lb \gamma\Rb\,\tilde Q_{n}^1 \,e^{ -\tilde\Delta_n\,\Y} \right]\nonumber \\
 &\simeq&
\Y\sum_{n=1}^{n_0}  \bar\beta_n\, e^{ -\tilde\Delta_n\,\Y}\,+\,\sum_{n=1}^{n_0}  \beta_n\, e^{ -\tilde\Delta_n\,\Y}\,+\,e^{-Y}\,\sum^\infty_{n=n_0+1}e^{-\gamma\,n}\nonumber \\
&=&
\Y\sum_{n=1}^{n_0}  \bar\beta_n\, e^{ -\tilde\Delta_n\,\Y}\,+\,\sum_{n=1}^{n_0}  \beta_n\, e^{ -\tilde\Delta_n\,\Y}\,+\,e^{-\Y}\,
{e^{-\gamma\,n_0}\over 1-e^{-\gamma}}\end{eqnarray}
The expression above simultaneously incorporates both the initial condition and large $Y$ asymptotics. 
So far, the only undetermined entry in \eq{beta} and the $S$-matrix is $\alpha^\prime_n$. 
It is not  trivial to compute $\alpha^\prime_n$ because it implies differentiation of $\alpha_n$
over discrete index $n$, requiring careful analytical continuation in $n$. 
 The expression for $\alpha^\prime_n$ is derived in the Appendix \ref{alphaprime}: we first relate all $\alpha^\prime_n$
to $\alpha^\prime_1$ (see (\ref {M201})) and then provide expression for $\alpha^\prime_1$
(see (\ref{E802}) for representation in the form of an infinite series and 
(\ref{E801}) for an alternative, integral representation).

In the limit of small $\gamma$, i.e. 
     $\gamma\,n\,<1$, the expressions are
\beq\label{alphaprime}
\alpha_n \simeq {1\over (n-1)!}~~~~~~~~~\rightarrow ~~~~~~~ \alpha^\prime_n \simeq {\gamma_E\,-\,H_{n-1} \over (n-1)!};~~~~~~~~~~\alpha^\prime_1 \simeq \gamma_E;
\eeq
 \begin{eqnarray}
\bar\beta_n =   {e^{\gamma\,n^2/4 \,-\, 9\,\gamma\,n/4}\over \ga^{n-1}\, (n-1)!}\,; ~~~~~~~~
\beta_n=    {e^{\gamma\,n^2/4 \,-\, 5\,\gamma\,n/4}\over \ga^{n}\, (n-1)!}\,\left[ {\ln(\gamma)}\, \,
- (n/2\,-\,5/4) \,\gamma\, -\gamma_E\,+\,H_{n-1} \right]\end{eqnarray}
 As mentioned earlier,  the leading large $\tilde Y$ asymptotics comes from $n=1$, reproducing  \eq{SA51}. Furthermore,
 $ S\Lb \Y=0\Rb\,=\,\sum_{n} \beta_n $. 
 
 Although we have not proven rigorously that the contribution of the circle at infinity is negligible, and the two contours on Fig. \ref{contr} can be deformed into each other, the fact that we found an expression for the $S$-matrix that yields correct both low and high rapidity limits is a strong indication that this is indeed the case. A more thorough investigation of this point is given in Appendix \ref{App A}.

\begin{boldmath}
\subsection{Scattering on  a target nucleus}
\end{boldmath}

We now consider  a  large nucleus target, modeling  nucleus in its rest frame as $A$ dipoles.  
This is   typical  dilute-on-dense process, in QCD frequently  described by the
BK equation \cite{BK}. In the BFKL cascade model of Section 2, the $S$-matrix computed 
 in  the nucleus  rest frame is
\beq \label{SAA02}
S^{BK}\Lb Y \Rb=\sum_{n=1}^{\infty} e^{ - \gamma\,n\,A}P^{\mbox{\tiny  BFKL}}_n\Lb \Y\Rb =\sum_{n=1}^{\infty} e^{ - \gamma\,n\,A} \frac{1}{N(Y) \,-\,1}\Bigg( 1 \,-\,\frac{1}{N(Y)}\Bigg)^n\,\,=\,\,\frac{ 1 }{1\,\,-\,\,\Big( 1\,-\,e^{ \gamma\,A}\Big) N(Y)}
\eeq
which indeeds satisfies  the "BK equation" \cite{BK}
\beq \label{SAA03}
\frac{ d\,S^{BK}\Lb Y\Rb}{d\,Y}\,\,=\,\,\Delta\Bigg((S^{BK})^2\,\,-\,\,S^{BK}\Bigg)
\eeq
The large $Y$ asymptotics 
\beq
S^{BK}\Lb Y\rightarrow\infty \Rb\simeq \frac{ e^{-\Delta\,Y} }{e^{  \,\gamma\,A}\,-\,1}
\xrightarrow{A\rightarrow\infty} e^{-\gamma\,A}\,e^{- \Delta\,Y}
\eeq
while at small rapidities we have  multi-Pomeron expansion:
\beq
S^{BK}\Lb Y\rightarrow 0 \Rb= 1\,+\,\Big( 1\,-\,e^{ \gamma\,A}\Big)\, e^{\Delta\,Y}\,+\,
\Big( 1\,-\,e^{ \gamma\,A}\Big)^2\, e^{2\,\Delta\,Y}\,+\,\cdots
\xrightarrow{A\rightarrow\infty} 1\,-\,e^{ \gamma\,A}\, e^{\Delta\,Y}\,+\,
e^{2 \gamma\,A}\, e^{2\,\Delta\,Y}\,+\,\cdots
\eeq
As we have extensively discussed above, the BFKL cascade does not  satisfy the $t$-channel unitarity (independence of the calculational frame). Therefore,  the $S$-matrix computed in the UTM will differ from  \eq{SAA02}. 

In  the UTM, the  $S$-matrix  
\beq \label{SAA1}
S^A\Lb \Y \Rb \, =\,\sum_{n=1}^{\infty} e^{ - \gamma\,n\,A} \,P^{\mbox{\tiny  UTM}}_n\Lb \Y\Rb\eeq
Equivalently,
\begin{eqnarray} \label{MISHA3N}
S^A\Lb   \Y \Rb\,&=&{1\over 2\pi i}\
\oint_{C} {\pi\,{(-1)}^ndn\over \sin(\pi \,n)}\, C_n^{(1)}\,\Phi_n\Lb u=e^{-\gamma\,A} \Rb e^{ \Delta_n\,\Y} \nn \\ 
&=&-\, \sum^A_{n=1}\,C_{-n}^{(1)}\,\,\Psi_n\Lb u=e^{-\gamma\,A} \Rb e^{- \tilde\Delta_n\,\Y} + {1\over\gamma}\,\sum^\infty_{n=A+1} {d\over dn} \left[C^{(1)}_{-n}\,\tilde Q_{n}^A \,e^{ -\tilde\Delta_n\,\Y} \right]\
\end{eqnarray}
We notice that for $A>n$, $\Psi_n$ is not singular. Hence  the leading large $Y$ asymptotics   given by $n=1$ is 
very easy to write down:
\beq
S^A\Lb   \Y \rightarrow\infty \Rb\,\simeq\, {e^{- \tilde\Delta_1\,\Y}\over e^{\gamma\,A}\,-\,e^{\gamma}}\,
\xrightarrow{A\rightarrow\infty} e^{-\gamma\,A}\,e^{- \tilde\Delta_1\,\Y}
\eeq
Pomeron expansion (the first line in \eq{MISHA3N})  is more suitable for small rapidities:
\begin{eqnarray}\label{Aexpan}
S^A\Lb   \Y \rightarrow 0 \Rb&=&1+ e^\gamma\, (1- e^{\gamma\,A})\,e^{ \Delta_1\,\Y}\,+\,
e^{2\,\gamma} (1- e^{\gamma\,A}) (1 - e^{\gamma\,(A+1)}) \,e^{ \Delta_2\,\Y}+\cdots\nn \\
&&\xrightarrow{A\rightarrow\infty} 1- e^\gamma e^{\gamma\,A}\,e^{ \Delta_1\,\Y}\,+\,
e^{2\,\gamma}  e^{2\gamma\,A} \,e^{ \Delta_2\,\Y}\,+\cdots
\end{eqnarray}
Comparing the asymptotics of the  BFKL and UTM  models, we notice that for small $\gamma$ and moderate rapidities they are practically identical. 
In particular the $A$ dependence in both models is the same for $A>1/\gamma$. 

\begin{boldmath}
\section{Summary}
\end{boldmath}

In this paper we have further explored the toy world focussing  on the BFKL cascade model and the UTM that  has been devised  to satisfied the $t$-channel unitarity, that is independence of the $S$-matrix on the reference frame.  

We have solved for the eigenfunctions and spectrum of both models. The $S$-matrix for high energy scattering 
is given in terms of the probabilistic formula \eq{SMS}. 
Both model Hamiltonians possess  negative and positive eigenvalues, each set separately  corresponding to a complete basis. 
The probabilities $P_n$ entering \eq{SMS} are naturally expanded in the basis of negative eigenvalues
with the expansion  truncated to a finite ($n$) number of terms, resulting in a well defined 
calculational  procedure both for the probabilities and for the $S$-matrix via    \eq{SMS}.

The Pomeron calculus emerges as an expansion of the $S$-matrix in terms of the eigenfunctions with the positive eigenvalues, 
the latter being identified with the dressed Pomeron intercepts.  In the BFKL cascade,
the intercept of $n$-Pomeron exchange scales linearly with $n$ as $n\,\Delta$, where $\Delta$ is the intercept of 
a single "BFKL Pomeron".  The expansion for the $S$-matrix is asymptotic, yet Borel summable.

In the UTM (see \eq{SA} for dipole-dipole scattering), the $n$-Pomeron intercept grows much faster with $n$, so that $\Delta_n > n\,\Delta_1$. 
In fact, at large $n$, the intercept grows exponentially
$\Delta_n \simeq \exp\Lb \gamma\,n\Rb $. Thus multi-Pomeron expansion in the UTM 
 cannot be summed a la Borel. Similar though somewhat less severe problem has been known in QCD for a long time: the $1/N_c$ corrections to the intercepts 
also lead to $\Delta_n^{QCD} > n\,\Delta_1^{BFKL}$
\footnote{A brief review of the problem, references and  the most pessimistic point of view on this problem can be found in Ref.\cite{LENC}.}. 
This lead  many experts (including  one of us \cite{LENC}) to believe  that the Pomeron calculus has a deadly problem
beyond strict leading order in $1/N_c$.

In the present paper we have managed to establish a resummation procedure that provides a meaningful $S$-matrix 
in the UTM, starting from the Pomeron calculus (see \eq{SA}). The main idea is based on analytical continuation which maps the positive spectrum into negative one. The approach is particularly suitable for computing asymptotic behaviour of the $S$-matrix  at large rapidities.  One of the interesting new observations is that the approach to unitarity features double poles in the complex angular momentum plane, with the leading asymptotics   $S(Y)\sim Y \,e^{-\Delta\,Y}$.

The UTM can be interpreted as a theory of interacting  bare ("BFKL") Pomerons. To  make this representation explicit we 
split the UTM Hamiltonian into that of the BFKL and self-interactions, ~ ${\cal H}_{UTM}\,=\,{\cal H}_{0}\,+\,{\cal H}_{I}$. The "BFKL" Pomeron has bare intercept $\Delta$. Self-interactions, which involve infinitely  many vertices both of the splitting and merging type, "dress" the bare Pomeron resulting in the UTM spectrum.  Since the bare Pomeron is supercritical ($\Delta>0$),
the dominant contributions to the "dressing" of the intercept are due to enhanced diagrams (small bare Pomeron loops). 
The solution of the UTM is a means of summing such diagrams, the lesson which can hopefully be taken to QCD.

We have seen in Section 2 that in the BFKL cascade model, due to its frame dependence,  the results  computed in the central mass frame differ from those obtained in the target rest frame.
For dipole-dipole collisions, the target rest frame result (BK solution) is quoted in \eq{SA1b}. Compared with the UTM (see \eq{Aexpan} for $A=1$), even at small $\gamma$, we observe a significant difference: e.g. the residues of  two-Pomeron exchanges differ 
by a factor two. This is not surprising. Theoretically it is of course well understood that the BFKL cascade model (a.k.a. the BK equation) is not applicable for dipole-dipole, or somewhat more general dilute-on-dilute scattering, beyond the leading single Pomeron exchange at small rapidities. It is interesting however to see explicitly how the unitarization corrections not included in the BK, affect the scattering amplitude.
The BK model is devised rather to describe 
dilute-on-dense scattering. Modeling the dense system as $A$ target dipoles,
we see indeed that the BK solution (\ref{SAA02})  and
the UTM result coincide  as long as 
$\gamma$ is small,   and $\gamma A> 1$ as long as the rapidity is not too large. This is the standard condition for absence of Pomeron loops and dominance of  semi-enhanced (fan) diagrams. For large enough rapidities $\tilde Y>\frac{1}{\gamma^2}$ the two results diverge as the rate of the exponential decrease of the scattering matrix in the two models is slightly different even at small $\gamma$. 
 
 The existence of the set of  negative eigenvalues and the corresponding eigenfunctions in either the BFKL cascade or UTM is not a widely recognized feature. While positive eigenvalues correspond to intercepts of multi-Pomeron exchanges and is a convenient basis for expanding the $S$-matrix at low rapidities, the negative eigenvalues have a very different meaning. They dominate the approach of the $S$-matrix to. saturation at $Y\rightarrow\infty$ \footnote{Some discussion along these lines could be found  in \cite{Kovner:2006vb}. In 
  \cite{Kovner:2006vb} though the spectrum was deemed to be double degenerate, which is not what we have discovered in the toy models.  
 }. The analogous exponential approach to saturation has been observed in QCD, the so called Levin-Tuchin law \cite{LETU}. Although the BFKL cascade model (a.k.a. the BK equation) in QCD is certainly more complicated due to nontrivial transverse dynamics, we believe that one leading Levin-Tuchin exponent is the direct analog of the largest negative eigenvalue $-\tilde\Delta_1$ in our toy model. Additional negative eigenvalues (and corresponding eigenfunctions) should also exist in QCD  and their study is an interesting open question. We also note that a model analogous of the UTM has been also formulated in two transverse dimensions. This is the so called 'Diamond action" model \cite{diamond}, which has been shown to be $t$-channel unitary, although its $s$-channel unitarity has not been formally established.

We conclude by repeating  our hope that the lessons learnt from the toy world can be helpful in further development  of the BFKL Pomeron calculus and better understanding of the saturation dynamics in real QCD.
 
~

~

       {\bf Acknowledgements} 
     
   We thank our colleagues at Tel Aviv University  for
 encouraging discussions. 
 We also thank Carlos Contreras for useful discussions related to the subject of this work. This research    was supported  by 
  Binational Science Foundation  grant \#2022132. AK is supported by the NSF Nuclear Theory grant 2208387.
 ML's work was funded by Binational Science Foundation grant \#2021789  and by the ISF grant \#910/23. This research was also supported by
 MSCA RISE 823947 “Heavy ion collisions: collectivity and precision in saturation physics” (HIEIC).

 \appendix

\begin{boldmath}
\section{ Large $n$ behavior of the Pomeron exchanges }
\label{App A}
\end{boldmath}
As has been discussed, the scattering matrix can be written as the sum over 
different numbers of the Pomeron exchanges, viz.:
\beq \label{SM}
S\Lb \Y\Rb\,\,=\,\,Z\Lb e^{- \gamma},\Y\Rb \,\,\,\,
\,=\,\,\,\sum^\infty_{n=0} C_n^{(1)}\,\Phi_n\Lb e^{- \gamma},\gamma\Rb\, e^{ \Delta_n\,\Y}\
\eeq 
 where $e^{ \Delta_n\,\Y}$ is the Green's function of the exchanges of $n$ Pomerons.  The asymptotic series of \eq{SM} has two problems: the severe problem of large $n$ behaviour for $\Delta_n  \xrightarrow{n \gg 1} \,e^{\gamma\,n}$ and the  Gaussian behaviour of $C_n^{(1)}  = \exp\Lb \h \gamma n (n+1)\Rb$. The first problem has been solved, introducing $j$ summation in \eq{EXP}. It turns out that we can solve the second problem by replacinng
 \beq \label{SM2}
 C_n^{(1)}  = \exp\Lb \h \gamma n (n+1)\Rb\,\,=\,\, \intl_{-\infty }^{\infty } d\lambda\frac{e^{-\frac{(\gamma -2 \lambda )^2}{8 \gamma }} }{\sqrt{2 \pi  \gamma }} e^{\lambda\,n}
 \eeq
 Integral over $\lambda$ is well convergent. Using \eq{EXP} and \eq{SM2} we can rewrite $S$ - matrix as
  \beq \label{SM3} 
 S\Lb \Y\Rb\,\,=\,\,Z\Lb e^{- \gamma},\Y\Rb \,\,\,\,
\,=\,\,\,e^{ - \Y} \sum_{j=0} \frac{\Y^j}{j!} \sum^\infty_{n=0} \underbrace{\intl_{-\infty }^{\infty } d\lambda\frac{e^{-\frac{(\gamma -2 \lambda )^2}{8 \gamma }} }{\sqrt{2 \pi  \gamma }} e^{\lambda \,n}}_{ C_n^{(1)}} e^{ \gamma\,n\,j} \Phi_n\Lb e^{- \gamma},\gamma\Rb
\eeq
At large $n$, $ \Phi_n\Lb e^{- \gamma},\gamma\Rb \,\to\,\Lb - e^\gamma\Rb^n$ and therefore, \eq{SM3} can be summed using the Borel transform:
\bea\label{SM4}
S\Lb \Y\Rb\,\,
\,&=&\,\,\,e^{ - \Y} \sum_{j=0} \frac{\Y^j}{j!} \intl_{-\infty }^{\infty } d\lambda\frac{e^{-\frac{(\gamma -2 \lambda )^2}{8 \gamma }} }{\sqrt{2 \pi  \gamma }} \int^\infty_0 d \tau e^{- \tau} \sum^\infty_{n=1} e^{ \Lb \lambda + \gamma\,j\Rb\,n} \tau^n \frac{ \Phi_n\Lb e^{- \gamma},\gamma\Rb}{n!}\nn\\
,&=&\,\,\,e^{ - \Y} \sum_{j=0} \frac{\Y^j}{j!} \intl_{-\infty }^{\infty } d\lambda\frac{e^{-\frac{(\gamma -2 \lambda )^2}{8 \gamma }} }{\sqrt{2 \pi  \gamma }} \int^\infty_0 d \tau\,e^{- \tau} \mathcal{B}\Lb \lambda, j ,\tau\Rb  \eea
where Borel function $ \mathcal{B}\Lb \lambda, j ,\tau\Rb$ is analytical function of all its variables, since the series in $n$ is the absolute converged series as well as sum over $j$ and integrals over $\lambda $ and $\tau$.

\section{Computing $\alpha^\prime_n$} \label{alphaprime}

In this Appendix we compute  $\alpha^\prime_n$ defined  in \eq{MISHA1}.  Notice the recurrent relation:
 \begin{subequations} \begin{eqnarray}
\alpha_{n+1}&=&\alpha_{n}\, {\gamma\over 2\sinh[\gamma n/2]} \label{M10} \\ \alpha^\prime_{n+1}&=&\alpha^\prime_{n}\, {\gamma\over 2\sinh[\gamma n/2]}\,-\,
\alpha_{n}\, {\gamma^2 \cosh[\gamma n/2]\over 4\sinh^2[\gamma n/2]}\,=\,
\alpha^\prime_{n}\, {\gamma\over 2\sinh[\gamma n/2]}\,-\,
\alpha_{n+1}\, {\gamma \cosh[\gamma n/2]\over 2\sinh[\gamma n/2]}\label{M20}
\end{eqnarray}
 \end{subequations} 
Thus we can relate all the derivatives of $\alpha_n$ to the one at $n=1$:
\beq \label{M201}
  \alpha^\prime_{n}=\alpha^\prime_1\, \alpha_n\,-\, \gamma\,
\alpha_n\, {1\over 2} \sum_{k=1}^{n-1}  \coth[\gamma k/2];~~~\frac{ \alpha^\prime_{n}}{ \alpha_n} = \alpha^\prime_1\, -\, \gamma\,
 {1\over 2} \sum_{k=1}^{n-1}  \coth[\gamma k/2] \eeq
Our next goal is to analytically continue $\alpha_n$ from integer $n$ to  continuous. From \eq{MISHA1},
\beq \label{E50}
\ln \alpha_n\,=\, - \ln[ (n-1)! ]  - \sum_{n=1}^{n-1}\ln\left[\frac{ 2 \sinh\Lb\h l \gamma\Rb}{\gamma l}\right] \,=\, - \ln[ (n-1)! ]  -  \sum_{k=1}^\infty\mu_k \,\gamma^{2 k}\, \sum^{n-1}_{l=1}  l^{2 k} \eeq
The last expression is an analytic function of $n$:
\beq \label{E602}
s_k(n)  \equiv \sum^{n-1}_{l=1} l^{2 k}  = H_{n-1}^{(-2 k)};~~~
\frac{d s_k(n)}{d n} = -2 k \left(\zeta (1-2 k)-H_{n-1}^{(1-2 k)}\right);
~~\frac{d s_k(n)}{d n}\Bigg{|}_{n=1} = -2 k \zeta (1-2 k);
\eeq
 $\alpha^\prime_1$ then has several equivalent representations:
\bea 
\hspace{-0.8cm}\alpha'_1 &=& \gamma_E - \sum^\infty_{k=1}\mu_k \gamma^{2 k}\Lb -2 k \zeta (1-2 k)\Rb \nonumber \\
&=& 
\gamma_E - 2\sum^\infty_{k=1} (-1)^k \left({\gamma\over 2\pi}\right)^{2 k}\,\zeta(2k) \,\zeta (1-2 k) {  = \gamma_E - 4\sum^\infty_{k=1}  \left({\gamma\over 4\pi^2}\right)^{2 k} \Gamma\Lb 2 k\Rb \zeta^2(2k) } \label{E802}\\
 &=&  \gamma_E - 4\sum^\infty_{k=1} \left({\gamma\over 4\pi^2}\right)^{2 k}\zeta(2 k) \int^\infty_0 dt \frac{t^{2k - 1}}{e^t - 1} 
=  {\color{black} \gamma_E - 2 \int^\infty_0 \frac{dt}{t (e^t - 1)} \Bigg\{ 1 - \frac{\gamma t}{4 \pi} \cot\Lb \frac{\gamma t}{4 \pi} \Rb\Bigg\}}\label{E801} \eea
 In the integral representation for $\alpha^\prime_1$, one  notices that $\cot$ has poles along the integration counter, so the integral should be understood in the principle value sense. For small $\gamma$,  $\alpha^\prime_1=\gamma_E$, while the first  correction is quadratic in $\gamma$  as is obvious from \eq{E801}.


\begin{thebibliography}{99}
     \frenchspacing
      
 
 
  \bibitem{ACJ}
  D.~Amati, L.~Caneschi and R.~Jengo,
  Nucl.\ Phys.\ B {\bf 101} (1975) 397.
  \bibitem{AAJ}
  V.~Alessandrini, D.~Amati and R.~Jengo,
  Nucl.\ Phys.\ B {\bf 108} (1976) 425.

  \bibitem{JEN}
   R.~Jengo,
  Nucl.\ Phys.\ B {\bf 108} (1976) 447.
    \bibitem{ABMC}
   D.~Amati, M.~Le Bellac, G.~Marchesini and M.~Ciafaloni,
  Nucl.\ Phys.\ B {\bf 112} (1976) 107.
\bibitem{CLR}
 M.~Ciafaloni, M.~Le Bellac and G.~C.~Rossi,
  Nucl.\ Phys.\ B {\bf 130} (1977) 388.
   
  \bibitem{CIAF}
    M.~Ciafaloni,
  Nucl.\ Phys.\ B {\bf 146} (1978) 427.
  
 
 \bibitem{MUSA}
A.~H.~Mueller and G.~P.~Salam,
  Nucl.\ Phys.\ B {\bf 475}, 293 (1996),  \\
  G.~P.~Salam,
  Nucl.\ Phys.\ B {\bf 461}, 512 (1996).
   [hep-ph/9509353].         
  
  \bibitem{RS}
  P.~Rembiesa and A.~M.~Stasto,
  Nucl.\ Phys.\ B {\bf 725} (2005) 251.
  [hep-ph/0503223].

 \bibitem{BIT}
   J.-P.~Blaizot, E.~Iancu and D.~N.~Triantafyllopoulos,
  Nucl.\ Phys.\ A {\bf 784} (2007) 227.
  [hep-ph/0606253].
   
  \bibitem{SHXI}  
    A.~I.~Shoshi and B.~W.~Xiao,
  Phys.\ Rev.\ D {\bf 73} (2006) 094014.
  [hep-ph/0512206].
  \bibitem{KOLEV}
   M.~Kozlov and E.~Levin,
  Nucl.\ Phys.\ A {\bf 779} (2006) 142.
  [hep-ph/0604039].
   
  \bibitem{KLremark2} 
  A.~Kovner and M.~Lublinsky,
  Nucl.\ Phys.\ A {\bf 767} 171 (2006).
  [hep-ph/0510047].
  
  	
\bibitem{nestor} N. Armesto, S. Bondarenko, J. G. Milhano and P. Quiroga, JHEP 0805 (2008) 103.
  
\bibitem{LEPRI}
  E.~Levin and A.~Prygarin,
  Eur.\ Phys.\ J.\ C {\bf 53} (2008) 385.
  [hep-ph/0701178].
     \bibitem{utm}
A.~Kovner, E.~Levin and M.~Lublinsky,
JHEP \textbf{08} (2016), 031.
[arXiv:1605.03251 [hep-ph]].
      

   
   \bibitem{utmm} 
A.~Kovner, E.~Levin and M.~Lublinsky,
JHEP \textbf{05}, 019 (2022)
[arXiv:2201.01551 [hep-ph]].

\bibitem{MUDI}
  A.~H.~Mueller,
  Nucl.\ Phys.\ B {\bf 415} (1994) 373;\,\,\,
  Nucl.\ Phys.\ B {\bf 437} (1995) 107;
  [hep-ph/9408245];\\
      A.~H.~Mueller and B.~Patel, 
      Nucl. Phys. B 425, 471, 1994;  e-Print: hep-ph/9403256 [hep-ph]

      
 \bibitem{BFKL}
   V.~S. Fadin, E.~A. Kuraev and L.~N. Lipatov,
\newblock Phys. Lett. {\bf B60}, 50 (1975);\,\,\,
E.~A. Kuraev, L.~N. Lipatov and V.~S. Fadin,
\newblock Sov. Phys. JETP {\bf 45}, 199 (1977),
\newblock [Zh. Eksp. Teor. Fiz.72,377(1977)];\,\,\,
I.~I. Balitsky and L.~N. Lipatov,
\newblock Sov. J. Nucl. Phys. {\bf 28}, 822 (1978),
\newblock [Yad. Fiz.28,1597(1978)].
  
\bibitem{BK}
I.~Balitsky,
{Phys.\ Rev.} {\bf D60}, 014020 (1999);
[arXiv:hep-ph/9812311];\,\,\,\,
Y.~V.~Kovchegov,
{Phys.\ Rev.}  {\bf D60}, 034008  (1999).
[arXiv:hep-ph/9901281].


\bibitem{BKP}
J.~Bartels,
Nucl. Phys. B \textbf{175} (1980), 365-401
J.~Kwiecinski and M.~Praszalowicz,
Phys. Lett. B \textbf{94} (1980), 413-416
   



      \bibitem{RY}
I. Gradstein and I. Ryzhik, {\it `` Table of Integrals, Series, and Products''},
Fifth Edition, Academic Press, London, 1994.
  
   \bibitem{LELU}
   E.~Levin and M.~Lublinsky,
Nucl. Phys. A \textbf{730} (2004), 191-211,
[arXiv:hep-ph/0308279 [hep-ph]];
  Phys.\ Lett.\ B {\bf 607} (2005) 131;
       [hep-ph/0411121];\,\,\,
  Nucl.\ Phys.\ A {\bf 763} (2005) 172,  
   [hep-ph/0501173].
  \bibitem{LETU}
E.~Levin and K.~Tuchin,
  Nucl.\ Phys.\ B {\bf 573}, 833 (2000);
  [hep-ph/9908317];\,\,\,
  Nucl.\ Phys.\ A {\bf 691}, 779 (2001);
  [hep-ph/0012167]; 
   Nucl.\ Phys.\ A   {\bf 693}, 787 (2001)
  [hep-ph/0101275].
  \bibitem{KOLEB}
Yuri V. Kovchegov and Eugene Levin, {\it `` Quantum Chromodynamics at High Energies"}, Cambridge Monographs on Particle Physics, Nuclear Physics and Cosmology, Cambridge University Press, 2012 .  
 
    
   
  
\bibitem{BORSUM}
 	Bruce Shawyer and Bruce Watson, {\it ``Borel's Method of Summability, Theory and Application''} , Clarendon Press, Oxford,1994;\\
	Ovidiu Costin,{``Asymptotocs and Borel Summability''}, Chapman \& HALL/CRC Monographs and Surveys in Pure and Applied Mathematics, CRC Press, Taylor \& Frencis Group,2009;\\$http://www1.phys.vt.edu/~ersharpe/spec-fn/app-d.pdf$. 
  \bibitem{LENC}
 E.~Levin,
Eur. Phys. J. C \textbf{83} (2023) no.5, 452.
[arXiv:2301.09337 [hep-ph]].
	\bibitem{AS}
  M. Abramowitz and I. Stegun, {\it ``Handbook of Mathematical Functions with Formulas, Graphs, and Mathematical Tables''}, United States Department of Commerce, National Bureau of Standards,1964.
\bibitem{QP}
 $https://en.wikipedia.org/wiki/Q-Pochhammer_-  symbol $  
 

      \bibitem{Kovner:2006vb}
A.~Kovner and M.~Lublinsky,
Nucl. Phys. A \textbf{779} (2006), 220-243; e-Print: hep-ph/0604085 [hep-ph]
	  
 \bibitem{diamond} A. Kovner, E. Levin, Ming Li and M. Lublinsky, JHEP 10 (2020) 185.
  e-Print: 2007.12132 [hep-ph]
	 \end{thebibliography}
\end{document}